\documentclass[twocolumn,epjc3]{svjour3}

\usepackage[T1]{fontenc}
\usepackage[utf8]{inputenc}
\usepackage{color}
\usepackage{graphicx}
\usepackage{amsmath}
\usepackage{amsfonts}
\usepackage{mathdots}
\usepackage{mathrsfs} 
\usepackage{dsfont} 
\usepackage{stmaryrd}
\usepackage{enumerate}
\usepackage{hyperref}
\hypersetup{colorlinks=true,
            linkcolor=blue,
            citecolor=blue,
            filecolor=green,
            urlcolor=cyan 
}

\usepackage[numbers,sort&compress]{natbib}
\usepackage{scalefnt}
\usepackage{txfonts}
\usepackage{microtype}

\journalname{Eur. Phys. J. A}
\renewcommand{\email}[1]{e-mail: \href{mailto:#1}{#1}}

\newcommand{\bra}[1]{\langle #1 \vert}
\newcommand{\ket}[1]{\vert #1 \rangle}
\newcommand{\scal}[2]{\langle #1 \vert #2 \rangle}
\newcommand{\elma}[3]{\bra{#1} #2 \ket{#3}}

\newcommand{\etal}{\emph{et al.}}
\newcommand{\abinitio}{\emph{ab initio~}}

\newcommand{\iunit}{\text{i}}
\newcommand{\TAURUS}{\textsf{TAURUS}}
\newcommand{\TAURUSvap}{$\textsf{TAURUS}_{\textsf{vap}}$}
\newcommand{\TAURUSpav}{$\textsf{TAURUS}_{\textsf{pav}}$}

\graphicspath{{.}{figures/}}

\begin{document}

\title{Symmetry-projected variational calculations with the numerical suite \TAURUS}
\subtitle{I. Variation after particle-number projection}

\author{B.~Bally\thanksref{ad:dft,em:bb} \and A.~Sánchez-Fernández\thanksref{ad:dft,em:asf} \and T.~R.~Rodríguez\thanksref{ad:dft,ad:ciaff,em:trr}}
\date{Received: \today{} / Accepted: date}

\thankstext{em:bb}{\email{benjamin.bally@uam.es}}
\thankstext{em:asf}{\email{adrian.sanchezf@estudiante.uam.es}}
\thankstext{em:trr}{\email{tomas.rodriguez@uam.es}}

\institute{
\label{ad:dft}
Departamento de F\'isica Te\'orica, Universidad Aut\'onoma de Madrid, E-28049 Madrid, Spain  
\and 
\label{ad:ciaff}
Centro de Investigaci\'on Avanzada en F\'isica Fundamental-CIAFF-UAM, E-28049 Madrid, Spain 
}

\maketitle
%
%
\begin{abstract}
We present the numerical code \TAURUSvap~that solves the variation after particle-number projection equations
for real general Bogoliubov quasiparticle states represented in a spherical harmonic oscillator basis.
The model space considered is invariant under spatial and isospin rotations but no specific 
set of orbits is assumed such that the code can carry out both valence-space and no-core calculations. 
In addition, no number parity is assumed for the Bogoliubov quasiparticle states such that the code can be used to 
describe even-even, odd-even and odd-odd nuclei. 
The variational procedure can be performed under several simultaneous constraints on the expectation values of a variety of 
operators such as the multipole deformations, the pairing field or the components of the angular momentum. 
To demonstrate the potential and versatility of the code, we perform several example calculations 
using an empirical shell-model interaction as well as a chiral interaction.
The ability to perform advanced variational Bogoliubov calculations offered by this code will, we hope, 
be beneficial to the shell model and \abinitio communities.
\end{abstract}
%
%
\section*{PROGRAM SUMMARY}
\begin{description}[font=\normalfont\itshape]
  \item[Program title:] \TAURUSvap
  \item[License:] GNU General Public License version 3 or later
  \item[Programming language:] Fortran 2008
  \item[DOI:] \href{https://doi.org/10.5281/zenodo.4381279}{10.5281/zenodo.4381279}
  \item[Repository:] \href{https://github.com/project-taurus/taurus\_vap}{github.com/project-taurus/taurus\_vap} 
  \item[Nature of problem:]
   The generation of variationally optimized Bogoliubov quasiparticle states is the starting point of various
   many-body methods. Nevertheless, no sophisticated and publicly available numerical code exists that can 
   handle general Hamiltonians as constructed for example in state-of-the-art effective field theories.
  \item[Solution method:]
   We propose an efficient numerical code that can perform the variation after particle-number projection of real general
   Bogoliubov quasiparticle states in the spherical harmonic oscillator basis. 
   The code offers several degrees of parallelization and make use of fast mathematical routines from the \textsf{BLAS} and \textsf{LAPACK} libraries.
  \item[Additional comments:]
  The code \TAURUSvap~is the first of a numerical suite that permits state-of-the-art symmetry projected calculations and that will be
    published in full in the future.
\end{description}
%
%
\section{Introduction}
\label{sec:intro}

Variational methods constitute a cornerstone of quantum many-body methods in general and nuclear theory in particular. 
Their fundamental principle is to build the best possible approximations to the Hamiltonian eigenstates 
by determining the states that yield the lowest expectation values for the Hamiltonian 
within a predetermined set of trial wave functions. 
There are sophisticated variational techniques that look for a quasi-exact representation of the Hamiltonian eigenstates,
such as the Variational Monte Carlo methods \cite{Carlson15a}, but their high computational cost limits their application 
to the lightest nuclei. In order to be able to address heavier nuclei, and in a systematic fashion, it is necessary to consider
far more approximate variational ansätze.
Many variational schemes in nuclear physics explore the manifold of Bogoliubov quasiparticle states 
by solving self-consistently the Hartree-Fock-Bogoliubov (HFB) equations \cite{RS80a}.
These states have the double advantage of possessing a simple product state structure, making them computationally 
easy to handle, and of naturally incorporating pairing correlations, which are needed 
as most nuclei exhibit a superfluid character \cite{Bohr58a,Belyaev58a,RS80a}.
Bogoliubov quasiparticle states, however, in general are not eigenstates of the proton- or neutron-number operators and
therefore are not invariant under a gauge rotation in the Fock space \cite{BR86a}. 
Consequently, they break one of the fundamental symmetries of the nuclear Hamiltonian.
Although such a symmetry-breaking scheme is \emph{a priori} undesirable, it was recognized long ago that this
could be advantageous to grasp further correlation energy while conserving the product states as simple variational ansätze.
Nevertheless, to avoid the resulting symmetry dilemma pointed out by Löwdin \cite{Lykos63a}, the symmetries broken during the variational minimization have
to be eventually restored \cite{Peierls57a,Lowdin67a,MacDonald70a,Bally20a}.

The concepts of symmetry breaking and restoration have been used for decades in energy density functional (EDF)
calculations \cite{Bender03a,Egido16a,Robledo18a,Sheikh19a}. Nowadays, state-of-the-art multi-reference EDF (MR EDF) 
schemes include the breaking and restoration of several symmetries concurrently.
For example, in addition to the breaking of the gauge invariance, associated with a good number 
of particles, it is customary to break the rotational invariance, associated with a good total angular momentum, to include 
deformation correlations \cite{Bender08a,Bally14a,Borrajo15a}. Globally, the most important contributions come from the 
quadrupole deformations \cite{Bender06a,Rodriguez15a}.
Nevertheless, the inclusion of octupole deformations, which requires the use of parity breaking reference states, 
become crucial to understand the structure of certain isotopes \cite{Robledo11b,Bernard16a}.
Over the years, the physicists working within the EDF community have developed many sophisticated numerical codes that can solve the HFB
equations considering a variety of constraints or intrinsic symmetries for the Bogoliubov reference states. 
Several of them have been published and are freely available \cite{Bennaceur05a,Carlsson10a,Maruhn14a,Ryssens15a,Perez17a,Schunck17a}. 
It is interesting to remark that public beyond-mean-field codes are far more scarce \cite{Schunck17a,Regnier18a}.
While being powerful tools, all those codes have the common feature that they were written and optimized 
specifically for energy density functionals and can not handle natively general Hamiltonians such as the ones 
constructed from Effective Field Theory (EFT) \cite{Epelbaum09a,Machleidt11a,Entem03a,Ekstrom15a,Soma20a} or 
commonly used in shell-model calculations \cite{Caurier05a,Poves01a,Brown06a,Nowacki09a}.
On that note, we remark that a valence-space finite-temperature HFB solver was recently proposed \cite{Ryssens20a}.

Nuclear \abinitio methods have experienced great progress in recent years and nowadays
routinely describe nuclei in the mid-mass region and even some heavy spherical nuclei \cite{Morris18a,Arthuis20a}.
This progress is in large part due to the development of a new generation of many-body methods that scale more gently
with the number of nucleons or the size of the model space considered \cite{Soma13a,Hagen14a,Hergert16a,Stroberg19a,Tichai20a}
than the quasi-exact methods previously used in very light systems \cite{Barrett13a,Carlson15a}. 
While the methods vary in their underlying principles, they usually rely either on a diagrammatic expansion on top of a reference state
or the diagonalization of the Hamiltonian within a set of reference states. In that regard, the choice of appropriate 
reference states, i.~e.~ many-body wave functions that capture the most important physical features of the problem, is 
critical to obtain a rapidly converging theory.
For example, the Slater determinant solution of the Hartree-Fock equations can be used to improve the convergence in Many-Body 
Perturbation Theory (MBPT) calculations \cite{Tichai16a}.
Not so surprisingly, the idea to use symmetry-breaking reference states has thus started to spread in the \abinitio 
community, eventually resulting in the first calculations of open-shell nuclei \cite{Soma13a,Hergert13a,Tichai18a,Yao20a}.

The interacting shell-model (ISM) \cite{Caurier05a} is another pillar in the modern description of nuclear structure. 
Diagonalizing exactly the many-body Hamiltonian within a restricted valence space, it offers to this day the most
accurate description of nuclear spectra not far away from shell closures. Recently, efforts have been carried out
to connect the ISM to the \abinitio project with the construction 
of valence-space effective interactions from first principles \cite{Stroberg19a,Holt19a}. 
Unfortunately, the combinatorial scaling of the ISM is, and will remain for the forseeable future, a major computational limitation. 
A way to overcome this difficulty might be to consider accurate approximations to the solutions of the Schrödinger equation 
such as the ones that can be constructed through Monte Carlo methods \cite{Otsuka01a,Abe12a}, 
or advanced multi-reference variational calculations in a valence space \cite{Schmid04a,Bally19a,Yao18a,Jiao18a}.

At the crossing point of these recent developments, we propose in this article the numerical code \TAURUSvap~that
 can solve the variation after  particle-number projection equations for real general Bogoliubov quasiparticle states represented
in a spherical harmonic oscillator basis.  This basis is a natural choice in many \abinitio or ISM calculations.
The code \TAURUSvap~presented here is actually the first of a new numerical suite, entitled \TAURUS, that is under development but was already
applied with success using both a shell-model interaction \cite{Bally19a} and a chiral interaction evolved through the In-Medium Similarity 
Renormalization Group (IMSRG) \cite{Yao20a}. The other codes of the suite will be the subject of later publications.

The article is organized as follows. 
In Sec.~\ref{sec:theory}, we give a detailed account of the theoretical framework.
Then, in Sec.~\ref{sec:examples}, we perform several example calculations using both an empirical shell-model interaction and a chiral Hamiltonian. 
In Sec.~\ref{sec:conclu}, we present a summary of this work as well as perspectives for the future developments of the numerical suite \TAURUS.
Finally, Sec.~\ref{sec:struct} and \ref{sec:detail} focus on the most relevant technical details of the program \TAURUSvap.

%
%
\section{Theoretical framework}
\label{sec:theory}

\subsection{Spherical harmonic oscillator}

First, let us consider the spherical harmonic oscillator (SHO) Hamiltonian
\begin{equation}
H_{\text{SHO}} = \frac{\mathbf{p}^2}{2m} + \frac12 m \omega^2 \mathbf{r}^2 \, ,
\end{equation}
where $\mathbf{p}$ is the momentum operator, $\mathbf{r}$ is the position operator, $\omega$ is the oscillator frequency, 
and $m$ is the average nucleon mass, i.e.\ $m = (m_p + m_n)/2$ with $m_p$ and $m_n$ being the proton and neutron 
masses, respectively. 
In addition, we define the oscillator length $b = \sqrt{\hbar / (m \omega)}$ with $\hbar$ being the reduced Planck constant.

\subsection{Single-particle basis}

We consider a model space spanned by a single-particle basis made of a set of eigenstates of $H_{\text{SHO}}$.
The reasons motivating this choice are manifold. 
First, this is the basis that is usually favoured in \abinitio and shell-model calculations such that we can
directly use the interactions constructed for those approaches and also easily benchmark our results.
Second, many expressions, such as the expectation values for the usual operators, are simple and therefore can be coded and debugged with ease.
Last but not least, it gives us the generality and flexibility to treat many problems on the same footing.
While for specific applications, it is of advantage to use a symmetry-adapted basis in order to reduce the computational cost or improve the convergence,
 e.g.\ to use an axial basis to compute axially deformed nuclei \cite{Perez17a}, our goal here is to build a versatile program. 

The SHO eigenstates $\phi_{n_a l_a s_a j_a m_{j_a} t_a m_{t_a}}$ are characterized by their principal quantum number $n_a$, 
orbital angular momentum $l_a$, spin $s_a = 1/2$, total angular momentum $j_a$ and its third component $m_{j_a}$, and isospin $t_a = 1/2$ and its third 
component $m_{t_a}$. We use as a convention $m_{t_a} = -1/2$ for proton single-particle states and $m_{t_a} = +1/2$ for neutron single-particle states.

For the sake of clarity, we use in the remainder of the article the shorthand notation 
\begin{equation*}
a \equiv (n_a , l_a, s_a , j_a, m_{j_a}, t_a, m_{t_a}).
\end{equation*}
The time-reversal partner of $a$ will be labeled 
\begin{equation*}
-a \equiv (n_a , l_a, s_a ,$ $j_a, -m_{j_a}, t_a, m_{t_a}). 
\end{equation*}
A multiplet containing all values of $\left\{ m_{j_a} = \right.$ $\left. -j_a , \ldots , j_a \right\}$ will be labeled  
\begin{equation*}
\hat{a} \equiv (n_a, l_a , j_a , s_a,t_a, m_{t_a}).
\end{equation*}
Finally, a multiplet containing all values of $\left\{ m_{j_a} = -j_a ,\right.$ $\left. \ldots , j_a \right\}$ and $\left\{ m_{t_a} = -1/2 , 1/2 \right\}$
will be labeled 
\begin{equation*}
\breve{a} \equiv (n_a, l_a , j_a , s_a , t_a).
\end{equation*}

The SHO eigenstates form an orthonormal set
\begin{equation}
 \scal{\phi_a}{\phi_b} = \delta_{ab} ,
\end{equation}
and they can be associated with a set of fermionic annihilation and creation operators $\left\{ c_a ; c^{\dagger}_a \right\}$.

Using the spherical coordinates $(\xi, \theta, \varphi)$, with $\xi = r/b$ being the dimensionless ratio between the radial distance $r$
and the oscillator length $b$, the explicit form of the SHO single-particle wave functions can be written as 
\begin{equation}
 \phi_{a} (\xi,\theta, \varphi) = R_{n_a l_a} (\xi) \Upsilon_{l_a s_a j_a m_{j_a}} (\theta, \varphi) \chi_{t_a m_{t_a}} \, ,
\end{equation}
where $R_{n_a l_a}$ is a normalized radial function, $\Upsilon_{l_a s_a j_a m_{j_a}}$ is a tensor spherical harmonic, and 
$\chi_{t_a m_{t_a}}$ is an isospin function.
The radial function $R_{n_a l_a}$ reads 
\begin{equation}
 R_{n_a l_a} (\xi) = \frac{\mathcal{N}_{n_a l_a}}{b^{3/2}} \xi^{l_a} e^{-\xi^2 / 2} L^{l_a + 1/2}_{n_a}(\xi^2) \, ,
\end{equation}
with $L^{l_a + 1/2}_{n_a}$ being a generalized Laguerre polynomial and $\mathcal{N}_{n_a l_a}$ the normalization constant
\begin{equation}
  \mathcal{N}_{n_a l_a} = \sqrt{\frac{2^{n_a + l_a + 2} n_a !}{\pi^{1/2} (2 n_a + 2 l_a + 1)!!}} \, .
\end{equation}
The tensor spherical harmonic $\Upsilon_{l_a s_a j_a m_{j_a}}$ reads
\begin{equation}
 \Upsilon_{l_a s_a j_a m_{j_a}} = \sum_{m_{l} = -l_a}^{l_a}  \sum_{m_{s} =-s_a}^{s_a} \langle l_a m_{l} s_a m_{s} | j_a m_{j_a} \rangle 
                                                                         Y_{l_a m_{l}} (\theta, \varphi) \chi_{s_a m_{s}} \, ,
\end{equation}
with $\langle l_a m_{l} s_a m_{s} | j_a m_{j_a} \rangle$ being a $SU(2)$ Clebsch-Gordan coefficient and $\chi_{s_a m_{s}}$ a spin function \cite{Varshalovich88a}.

In this work, we will only consider model spaces that are rotationally invariant, i.e.\ that includes all the members of a multiplet 
$\left\{ m_{j_a} = -j_a , \ldots , j_a \right\}$, and isospin invariant, i.e.\ with the same single-particle states for protons 
and neutrons: $\left\{ m_{t_a} = -1/2 , 1/2 \right\}$. Apart from that, no other limitation is imposed on the orbitals included. In particular,  we will consider both
no-core and valence-space model spaces.
With that, the model space is entirely specified by the set of orbits $\mathcal{M} \equiv \left\{ \breve{a} \right\}$ it contains
and its dimension can be calculated as 
\begin{equation}
 d_{\mathcal{M}} = 2 \sum_{\breve{a}}^{\mathcal{M}} (2j_a + 1) \, .
\end{equation}

Finally, let us comment that the single-particle basis is ordered such that the first half of the basis is made of the proton single-particle states while the second half
is made of the neutron single-particle states.

\subsection{Bogoliubov quasiparticle states}
\label{sec:bogo}

\subsubsection{Bogoliubov transformations}
We define the Bogoliubov state $\ket{\Phi}$ as the vacuum for a set of quasiparticle annihilation and creation operators 
$\left\{ \beta_k ; \beta^{\dagger}_k \right\}$ constructed through the unitary linear Bogoliubov transformation 
\begin{subequations}
\begin{equation}
\begin{pmatrix}
\beta \\
\beta^{\dagger}
\end{pmatrix}
=
\begin{pmatrix}
U^\dagger & V^\dagger \\
V^T & U^T
\end{pmatrix} 
\begin{pmatrix}
c \\
c^{\dagger}
\end{pmatrix}  
\equiv \mathcal{W}^{\dagger} 
\begin{pmatrix}
c \\
c^{\dagger}
\end{pmatrix}  \, , \label{bogophi} 
\end{equation}
with 
\begin{equation}
\mathcal{W} \mathcal{W}^\dagger = \mathcal{W}^\dagger \mathcal{W} = 1_{2d_{\mathcal{M}}} \, ,
\end{equation}
where $1_m$ refers to the identity matrix of dimensions $m \times m$.
\end{subequations}
The expanded form of Eq.~\eqref{bogophi} reads as
\begin{subequations}
\begin{align}
 \beta_{k} &= \sum_{i} U^*_{i k} \, c_{i} + V^*_{i k} \, c_{i}^\dagger \, ,  \\
 \beta_{k}^\dagger &= \sum_{i} U_{i k} \, c_{i}^\dagger + V_{i k} \, c_{i} \, .  
\end{align}
\end{subequations}

In the present work, we allow for general Bogoliubov transformations without any restriction except for the fact that the matrices ($U$,$V$)
are kept \emph{real}.\footnote{While we will consider only Bogoliubov matrices, and their related quantities, that are real, we will give the general complex definitions of most quantities.}
From a symmetry perspective, this means that, in the most general case discussed here,  we consider wave functions that break all symmetries but the complex conjugation.
In particular, angular-momentum, particle-number, and parity symmetry breaking, as well as proton-neutron mixing are allowed.
Bogoliubov transformations that conserve a certain quantum number $\mu$ are thus considered only as specific cases where the matrices ($U$,$V$) 
have particular block structures that do not mix SHO single-particles states with different value of $\mu$. Let us remark in particular that 
starting a calculation with such specific matrices ($U$,$V$), their structure will be preserved throughout the minimization procedure if the Hamiltonian 
and the constraints imposed during the minimization also conserve $\mu$ \cite{RS80a}.

\subsubsection{One-body densities}
\label{sec:bogo:dens}
A Bogoliubov state $\ket{\Phi}$ can be characterized by its one-body densities defined as
\begin{subequations}
\begin{align}
 \rho_{ij} &= \frac{\elma{\Phi}{c^\dagger_j c_i}{\Phi}}{\scal{\Phi}{\Phi}} = \left( V^* V^T \right)_{ij} , \label{eq:rho} \\ 
 \kappa_{ij} &= \frac{\elma{\Phi}{c_j c_i}{\Phi}}{\scal{\Phi}{\Phi}} =  \left( V^* U^T  \right)_{ij}     , \label{eq:kap} \\
 \kappa^*_{ij} &= \frac{\elma{\Phi}{c_i^\dagger c_j^\dagger}{\Phi}}{\scal{\Phi}{\Phi}}= \left( V U^\dagger \right)_{ij}   . \label{eq:kap*} 
\end{align}
\end{subequations}

Additionally, when dealing with two Bogoliubov states, e.g.\ $\ket{\Phi_L}$ with Bogoliubov matrices ($U_L$,$V_L$) and $\ket{\Phi_R}$ with Bogoliubov matrices ($U_R$,$V_R$), 
it is possible to define the so-called transition densities
\begin{subequations}
\begin{align}
 \rho^{LR}_{ij} &= \frac{\elma{\Phi_L}{c^\dagger_j c_i}{\Phi_R}}{\scal{\Phi_L}{\Phi_R}} = \left( V_R^* {A^T}^{-1}  V_L^T \right)_{ij} , \label{eq:rho0p} \\ 
 \kappa^{LR}_{ij} &= \frac{\elma{\Phi_L}{c_j c_i}{\Phi_R}}{\scal{\Phi_L}{\Phi_R}} = \left( V_R^* {A^T}^{-1} U_L^T  \right)_{ij}  , \label{eq:kap0p} \\ 
 {\kappa^{RL}_{ij}}^* &= \frac{\elma{\Phi_L}{c_i^\dagger c_j^\dagger}{\Phi_R}}{\scal{\Phi_L}{\Phi_R}} = \left( V_L {A}^{-1} U_R^\dagger \right)_{ij} , \label{eq:kapp0} 
\end{align}
\end{subequations}
where 
\begin{equation}
A = U_R^\dagger U_L + V_R^\dagger V_L  \, .
\end{equation}
Let us remark that in the diagonal case where $\ket{\Phi_L} = \ket{\Phi_R} = \ket{\Phi}$, we obtain $A = 1_{d_{\mathcal{M}}}$ due to the unitarity of $\mathcal{W}$ and, therefore,
the transition densities reduce to the usual densities defined above. 

\subsubsection{Quasiparticle excitations}

Starting from a Bogoliubov vacuum $\ket{\Phi}$, it is possible to create quasiparticle excitations by applying the quasiparticle creation operators on top of it. A $n$-quasiparticle excitation is defined as 
\begin{equation}
\ket{\Phi^{k_1 \ldots k_n}} \equiv \beta^\dagger_{k_1} \ldots  \beta^\dagger_{k_n} \ket{\Phi} .
\end{equation}
In practice, the quasiparticle excitations can be easily performed by interchanging the columns of the $U$ and $V$ matrices\footnote{Here, we assume that all the 
quasiparticle excitations are distinct.} \cite{RS80a}
\begin{equation}
 (U_{l k_{i}}, V_{l k_{i}}) \rightarrow (V^*_{l k_{i}}, U^*_{l k_{i}}) , \, \forall i \in \llbracket 1, n \rrbracket , \, \forall l \in \llbracket 1, d_\mathcal{M} \rrbracket .
\end{equation}
Note that starting from a Bogoliubov state with a number parity equal to $\pi_A$ (see Sec.~\ref{sec:pnp}), 
the state $\ket{\Phi^{k_1 \ldots k_n}}$ will have a number parity equal to $(-1)^n \pi_A$.

\subsection{Nuclear many-body Hamiltonian}

Our objective is to develop a versatile approach that can be applied to different type of model spaces and Hamiltonians.
Therefore, we will consider a general nuclear Hamiltonian that contains up to two-body terms
\begin{equation}
\begin{split}
H &= H^{(0)} + H^{(1)} + H^{(2)} \, , \\
  &= h^{(0)} + \sum_{ij} h^{(1)}_{ij} c_i^\dagger c_j + \frac{1}{(2!)^2} \sum_{ijkl} \overline{h}^{(2)}_{ijkl} c_i^\dagger c_j^\dagger c_l c_k \, ,
\end{split}
\end{equation}
where $h^{(0)}$, $h^{(1)}_{ij}$, and $\overline{h}^{(2)}_{ijkl}$ are zero-, one-, and antisymmetrized two-body matrix elements, respectively. 
In particular, we do not constrain the physical content of each term except to require that they respect the symmetries of $H$, such as the rotational 
and parity invariances, and are kept \emph{real}.
The matrix elements could be obtained from, for example, an empirical shell-model interaction, a normal-order chiral interaction or even 
a Hamiltonian-based energy functional.\footnote{As demonstrated in 
\cite{Dobaczewski07a,Lacroix09a,Bender09a,Duguet09a}, beyond-mean-field calculations based on general energy functionals are not well defined such
that we prefer not to consider them here.} 

While the inclusion of a three-body term would not present theoretical difficulties, the amount of computer memory required to store all the non-zero matrix 
elements in the SHO single-particle basis would be prohibitive in most model spaces. Also, \abinitio calculations have demonstrated that the most 
important contributions coming from a realistic 3N interactions can be included through the so-called normal order two-body approximation (NO2B) 
\cite{Roth12a,Gebrerufael16a}. Nevertheless, it might be desirable to include the full three-body term in small model spaces in the future. 

To illustrate our discussion, let us first consider the most typical form of an empirical shell-model interaction that is such that  
\begin{subequations}
\begin{align}
 h^{(0)} &= 0 \, , \\
 h^{(1)}_{ij} &= \delta_{ij} \epsilon_i \, , \\
 \overline{h}^{(2)}_{ijkl} &= \overline{V}^{(2)}_{ijkl} \, ,
\end{align}
\end{subequations}
where $\epsilon_i$ is a single-particle energy and $\overline{V}^{(2)}_{ijkl}$ a two-body matrix element obtained from a $G$-matrix and/or a fit.
We note that, although considered null here, $h^{(0)}$ could be taken as equal to the energy of the inert core considered for the valence space at hand.

Another relevant example is the case of a typical Hamiltonian constructed from EFT that can be decomposed as
\begin{equation}
H = T^{(1)} - T^{(1)}_{\text{com}} + V^{(2)} -  T^{(2)}_{\text{com}} + W^{(3)} \, ,
\end{equation}
where $T^{(1)}$ is the one-body kinetic energy, $V^{(2)}$ and $W^{(3)}$ are the 2N and 3N chiral interactions, respectively, 
and finally $T^{(1)}_{\text{com}}$ and $T^{(2)}_{\text{com}}$ are the one- and two-body corrections for the center-of-mass motion \cite{Hergert09a}, respectively. 
Considering the NO2B approximation of the three-body piece with respect to an arbitrary state \cite{Kutzelnigg97a}, we obtain the expressions
\begin{subequations}
\begin{align}
 h^{(0)} &= \tilde{w}^{(0)} \, , \\
 h^{(1)}_{ij} &= \left( 1-\frac{1}{A_0} \right) \elma{i}{\frac{\mathbf{p}^2_1}{2m}}{j} + \tilde{w}^{(1)}_{ij} \, , \\
 \overline{h}^{(2)}_{ijkl} &= \elma{ij}{V^{(2)} - \frac{\mathbf{p}_1\cdot\mathbf{p}_2}{mA_0}}{\overline{kl}} + \tilde{w}^{(2)}_{ijkl} \, , 
\end{align}
\end{subequations}
where $\ket{\overline{kl}} = \ket{kl} - \ket{lk}$, $A_0$ is the nucleon number of the nucleus we wish to describe, and finally, 
$\tilde{w}^{(0)}$, $\tilde{w}^{(1)}_{ij}$ and $\tilde{w}^{(2)}_{ijkl}$ are the matrix elements of the zero-, one- and two-body parts
of the normal-ordering of three-body interaction, respectively, whose expressions can be found for example in Ref.~\cite{Gebrerufael16a}.

\subsection{Particle-number projection}

\subsubsection{Basic principles}
\label{sec:pnp}
The proton-, neutron-, and nucleon-number operators are defined as
\begin{align}
Z &= \sum_{i} \delta_{m_{t_i} -1/2} \, c_i^\dagger c_i \, , \\
N &= \sum_{i} \delta_{m_{t_i} +1/2} \, c_i^\dagger c_i \, , \\
A &= Z + N \, .
\end{align}

For the sake of clarity, for the remainder of the section we will consider only the generic particle species $N$ for our definitions but, except told
otherwise, all the quantities have to be generalized by replacing $N$ by $Z$ or $A$.
With that said, the gauge rotations for the species $N$ in the Fock space are associated with the abelian group $U(1)_N$.
In this work, we consider its unitary representation
\begin{equation}
\left\{ R(\varphi_N) = e^{\iunit \varphi_N N} , \varphi_N \in [0,2\pi] \right\} \, ,
\end{equation}
with $\varphi_N$ and $R(\varphi_N)$ being the gauge angle and gauge rotation operator, respectively.
 Any eigenvector $\ket{\Theta^{N_0}}$ of the operator $N$ with eigenvalue
 $N_0$ represents a one-dimensional irreducible representation (irrep) of $U(1)_N$ such that
\begin{equation}
 R(\varphi_N) \ket{\Theta^{N_0}} = e^{\iunit \varphi_N N_0} \ket{\Theta^{N_0}} \, . 
\end{equation}
On the contrary, considering an arbitrary many-body wave function $\ket{\Theta}$ that may not be an eigenvector of $N$, it is possible
to use a projection operator 
\begin{equation}
\label{eq:PN}
P^{N_0} = \frac{1}{2\pi} \int_0^{2\pi} \! d \varphi_N \, e^{\iunit \varphi_N (N- {N_0})} \, ,
\end{equation}
such that 
\begin{equation}
N P^{N_0} \ket{\Theta} = N_0 P^{N_0} \ket{\Theta} \, ,
\end{equation}
and $P^{N_0} \ket{\Theta}$ is non-vanishing for at least one value of $N_0$.

In general, a Bogoliubov vacuum $\ket{\Phi}$ will not be an eigenstate for the particle species $N$. Only when the pairing correlations
vanish\footnote{As recently demonstrated, the zero-pairing limit of the HFB equations does not necessarily lead to the HF equations and has
to be handled with care \cite{Duguet20a,Duguet20b}.}
 and the Bogoliubov state reduces to a mere Slater determinant for this species, it is the case. 
On the other hand, the nuclear Hamiltonian $H$ is always invariant under gauge rotations for particle species $N$, i.e.\
\begin{equation}
 \left[ H, U(\varphi_N) \right] = 0 , \forall \varphi_N \in [0,2\pi]  .
\end{equation}
Therefore, to respect the symmetries of $H$, it is necessary to project $\ket{\Phi}$ onto the good number of particles of the nucleus
we wish to describe, for example $N_0$. 

Obviously, in the general case both the neutron and proton gauge invariances have to be restored, such that we construct the normalized 
projected state
\begin{equation}
 \ket{\Psi^{Z_0 N_0}} = \frac{P^{Z_0}P^{N_0}\ket{\phi}}{\sqrt{\elma{\Phi}{P^{Z_0}P^{N_0}}{\Phi}}} \, .
\end{equation}
It is important to stress that the projected state $\ket{\Psi^{Z_0 N_0}}$ thus obtained has generally a more complex structure
than the one of a mere product state. 

While a Bogoliubov state $\ket{\Phi}$ is not in general an eigenstate for any of the particle-number operators mentioned above, it is by construction
an eigenstate of the number-parity operator $\Pi_A = e^{\iunit \pi A}$ with an eigenvalue $\pi_A = +1$ or $-1$. 
As a consequence, the interval of integration of the projection operator for $A$ can be reduced to $[0,\pi]$
\begin{equation}
\label{eq:PA}
P^{A_0} = \frac{1}{\pi} \int_0^{\pi} \! d \varphi_A \, e^{\iunit \varphi_A (A- {A_0})} \, ,
\end{equation}
which is numerically more efficient \cite{Bally20a}. 
In addition, if there is no proton-neutron mixing in the Bogoliubov matrices ($U$,$V$), 
 the state $\ket{\Phi}$ will be an eigenstate of $\Pi_Z$ and $\Pi_N$ individually, thus  allowing for 
similar reductions for $P^{Z_0}$ and $P^{N_0}$.

In Sec.\ \ref{sec:discPNP}, we give the expressions for the discretized projection operators implemented in the code. More generally, 
further details on the projection on particle-number and its numerical implementation can be found for example in \cite{Bally20a,Sheikh19a}.

\subsubsection{Gauge-rotated quasiparticle states}
Because of the one-body character of gauge transformations, a gauge-rotated Bogoliubov quasiparticle state is still a Bogoliubov quasiparticle state.
Therefore, we can define the gauge-rotated state
\begin{equation}
 \ket{\Phi(\varphi)} \equiv R(\varphi_Z) R(\varphi_N) \ket{\Phi} ,
\end{equation}
where we used the shorthand notation $\varphi \equiv (\varphi_Z , \varphi_N)$ with $\ket{\Phi(0)} \equiv \ket{\Phi}$.
The Bogoliubov matrices ($U^\varphi$,$V^\varphi$) of the state $\ket{\Phi(\varphi)}$ can be expressed as 
\begin{subequations}
\begin{align}
 U^\varphi &= 
  \begin{pmatrix}
    e^{+ i \varphi_Z} 1_{d_{\mathcal{M}}/2} & 0 \\
    0 & e^{+i \varphi_N} 1_{d_{\mathcal{M}}/2}
  \end{pmatrix} U  , \\
 V^\varphi &=
  \begin{pmatrix}
    e^{- i \varphi_Z} 1_{d_{\mathcal{M}}/2} & 0 \\
    0 & e^{-i \varphi_N} 1_{d_{\mathcal{M}}/2}
  \end{pmatrix} V  ,
\end{align}
\end{subequations}
where we used our partition of the single-particle basis in terms of the proton (first half) and neutron (second half) single-particle states.

\subsubsection{Projected expectation values}
Let us consider an operator $O$ that is a scalar regarding gauge rotations, i.e.\
\begin{equation}
 \left[ O, U(\varphi_N) \right] = 0 , \forall \varphi_N \in [0,2\pi] .
\end{equation}
Its expectation value for the projected state $\ket{\Psi^{Z_0 N_0}}$ reads
\begin{equation}
 O^{Z_0 N_0} \equiv \elma{\Psi^{Z_0 N_0}}{O}{\Psi^{Z_0 N_0}} = \frac{\elma{\Phi}{O P^{Z_0}P^{N_0}}{\Phi}}{\elma{\Phi}{P^{Z_0}P^{N_0}}{\Phi}} \, .
\end{equation}
Expanding the numerator and denominator we obtain 
\begin{align}
  \elma{\Phi}{O P^{Z_0}P^{N_0}}{\Phi} &= \frac{1}{(2\pi)^2} \int_{0}^{2\pi} d\varphi_Z \int_{0}^{2\pi} d\varphi_N \, e^{-i (\varphi_Z Z_0 + \varphi_N N_0)} \nonumber \\
                                      &\phantom{=}  \times \frac{\elma{\Phi}{O}{\Phi(\varphi)}}{\scal{\Phi}{\Phi(\varphi)}} \scal{\Phi}{\Phi(\varphi)} ,
\end{align}
\begin{align}
  \elma{\Phi}{P^{Z_0}P^{N_0}}{\Phi} &= \frac{1}{(2\pi)^2} \int_{0}^{2\pi} d\varphi_Z  \int_{0}^{2\pi} d\varphi_N \, e^{-i (\varphi_Z Z_0 + \varphi_N N_0)} \nonumber \\
                                      &\phantom{=}  \times \scal{\Phi}{\Phi(\varphi)} .
\end{align}
In the above equations, the fundamental quantities that have to be evaluated are the norm overlap $\scal{\Phi}{\Phi(\varphi)}$ and the rotated kernel 
\begin{equation}
 O^{0 \varphi} \equiv \frac{\elma{\Phi}{O}{\Phi(\varphi)}}{\scal{\Phi}{\Phi(\varphi)}} . 
\end{equation}
The norm overlap is commonly calculated using the Pfaffian algebra \cite{Robledo09a,Avez12a} (see Sec.~\ref{sec:pfaf} for more information on the practical implementation).
Concerning the rotated kernel, it can be evaluated easily thanks to the Generalized Wick Theorem (GWT) \cite{Balian69a} when the overlap does not vanish. 
In particular, we remark that working within a Hamiltonian-based framework, we consider all the fully-contracted terms appearing in the application of the GWT.

Assuming now that $O$ contains up to two-body terms, i.e. \
\begin{equation}
\begin{split}
O &= O^{(0)} + O^{(1)} + O^{(2)} \, , \\
  &= o^{(0)} + \sum_{ij} o^{(1)}_{ij} c_i^\dagger c_j + \frac{1}{(2!)^2} \sum_{ijkl} \overline{o}^{(2)}_{ijkl} c_i^\dagger c_j^\dagger c_l c_k ,
\end{split}
\end{equation}
where $o^{(0)}$, $o^{(1)}_{ij}$, and $\overline{o}^{(2)}_{ijkl}$ are zero-, one-, and antisymmetrized two-body matrix elements, respectively, 
the rotated kernel reads 
\begin{equation}
\begin{split}
 \label{eq:expo}
 O^{0 \varphi} &=  o^{(0)} + \sum_{ij} o^{(1)}_{ij} \rho^{0\varphi}_{ji} + 
                 \frac{1}{2} \sum_{ijkl} \overline{o}^{(2)}_{ijkl}  \rho^{0\varphi}_{ki} \rho^{0\varphi}_{lj} \\ 
                 &\phantom{=} + \frac{1}{4} \sum_{ijkl}  \overline{o}^{(2)}_{ijkl} {\kappa^{\varphi 0}_{ij}}^* \kappa^{0\varphi}_{kl}  .
\end{split}
\end{equation}
where used the definitions of the transitions densities given in Sec.~\ref{sec:bogo:dens} setting $\ket{\Phi_L} = \ket{\Phi(0)}$
and $\ket{\Phi_R} = \ket{\Phi(\varphi)}$.
Let us remark in passing that the expectation value of a one-body operator can be obtained from the above equation 
by taking $\overline{o}^{(2)}_{ijkl} = 0$.

In the special case of the Hamiltonian, we define the transition fields
\begin{subequations}
\begin{align}
 \Gamma^{0\phi}_{ij} &= \sum_{kl} \overline{h}^{(2)}_{ikjl} \, \rho^{0\varphi}_{lk} , \\
 h^{0\phi}_{ij} &= h^{(1)}_{ij} + \Gamma^{0\phi}_{ij} , \\
 \Delta^{0\phi}_{ij} &= \frac12  \sum_{kl} \overline{h}^{(2)}_{ijkl} \, \kappa^{0\varphi}_{kl} , \label{eq:Dfield} \\
 {\Delta^{\phi 0}_{ij}}^* &= \frac12  \sum_{kl} \overline{h}^{(2)}_{klij} \, {\kappa^{\varphi 0}_{kl}}^* ,
\end{align}
\end{subequations}
such that the expression given in Eq.~\eqref{eq:expo} can be written as 
\begin{equation}
 \label{eq:exph}
 H^{0 \varphi} = h^{(0)} + \sum_{ij} h^{(1)}_{ij} \rho^{0\varphi}_{ji} + \frac{1}{2} \sum_{ij} \left( \Gamma^{0\phi}_{ij} \rho^{0\varphi}_{ji}
                  - \Delta^{0\phi}_{ij} {\kappa^{\varphi 0}_{ji}}^* \right) .
\end{equation}
The advantage of the above expressions is that they allow us to confine the use of the two-body matrix elements to the calculation
of the fields. Then, the fields can be used to compute the Hamiltonian rotated kernel and other quantities related to the projected gradient. 

\subsection{Variation after projection}
\label{sec:vap}

\subsubsection{Projected gradient}
The variational space explored in this work is the manifold of particle-number projected Bogoliubov quasiparticle states.
Therefore, we look for the solution of the variational equation
\begin{equation}
 \label{eq:VAP}
 \delta \elma{\Psi^{Z_0 N_0}}{H}{\Psi^{Z_0 N_0}} = 0 \, .
\end{equation}
This equivalent to looking for the Bogoliubov reference state $\ket{\Phi}$ that minimizes the projected
energy 
\begin{equation}
 E^{Z_0 N_0} \equiv \elma{\Psi^{Z_0 N_0}}{H}{\Psi^{Z_0 N_0}} = \frac{\elma{\Phi}{H P^{Z_0}P^{N_0}}{\Phi}}{\elma{\Phi}{P^{Z_0}P^{N_0}}{\Phi}} \, .
\end{equation}
This approach is called variation after particle-number projection (VAPNP) in the literature \cite{Anguiano01a,Anguiano02a,Stoitsov07a,Rodriguez05a} and is superior to the simpler
HFB scheme for several reasons. 
First, the trial ansatz considered is better on a fundamental level: it possesses a more general structure, being a linear superposition of infinitely many Bogoliubov vacua,
and is symmetry adapted, and therefore strictly explores a variational space with the correct number of particles.
In particular, by construction, the VAPNP can yield a lower energy than the particle-number projection after
variation (PNPAV) scheme within which the Bogoliubov state is obtained by solving the HFB equations first and is projected onto good particle numbers only afterwards.
In addition, the VAPNP gives a better description of pairing correlations \cite{Anguiano01a,Anguiano02a,Rodriguez05a,Romero19a}. Notably, it prevents the frequent and unphysical collapse
of the pairing often observed in HFB calculations \cite{Mang66a}. This fact has also a practical advantage in configuration mixing calculations as it reduces
the chance of finding a pair of orthogonal states, which forbids the use of the GWT usually employed in such calculations.

On the other hand, the VAPNP is more computationally costly than the PNPAV as the particle-number projection has to be performed at each iteration. 
Even using an efficient discretization of the gauge integrals, such as the one proposed by Fomenko \cite{Fomenko70a,Bally20a}, the numerical cost
generally increases by one or two order of magnitudes compared to traditional PNPAV or plain HFB calculations. Nevertheless, the computational power
nowadays at our disposal makes possible such calculations even in large model spaces \cite{Robledo18a}.

Solving Eq.~\eqref{eq:VAP} corresponds to finding the state $\ket{\Phi}$ such that the projected gradient \cite{Egido82a,RodriguezMAS}, with matrix elements
\begin{equation}
\begin{split}
 G_{ij} &= - \frac{\elma{\Phi}{ \beta_j \beta_i  (H - E^{Z_0 N_0}) P^{Z_0}P^{N_0}}{\Phi}}{\elma{\Phi}{P^{Z_0}P^{N_0}}{\Phi}} \\
        &\equiv - \left( H^{20,Z_0 N_0}_{ij} - E^{20,Z_0 N_0}_{ij} \right) ,
\end{split}
\end{equation}
vanishes
\begin{equation}
 || G ||_F = 0 \, ,
\end{equation}
where $||\cdot||_F$ stands for the Frobenius norm and the superscript $20$ indicates the part of the operator when expressed in the quasiparticle basis of $\ket{\Phi}$ \cite{RS80a}.

\subsubsection{Rotated gradient}
To compute the projected gradient, it is necessary to compute the values of the \emph{rotated} gradient at each gauge angle.
After a lengthy but straightforward derivation \cite{Anguiano01a,RodriguezMAS}, the matrix elements of the rotated gradient at angle $\varphi$ can be expressed as
\begin{align}
 G_{ij}^{0\varphi} &= \big( H^{0 \varphi}  - E^{Z_0 N_0} \big) R_{ij}^{\varphi}  + \left( \bar{U}^{\varphi} {\vphantom{\big(}}^T h^{0\phi} \bar{V}^{\varphi} \right)_{ij}
                      - \left( \bar{V}^{\varphi} {\vphantom{\big(}}^T h^{0\phi}{\vphantom{\big(}}^T \bar{U}^{\varphi} \right)_{ij} \nonumber \\
                      & + \left( \bar{U}^{\varphi} {\vphantom{\big(}}^T \Delta^{0 \phi} \bar{U}^{\varphi} \right)_{ij}
                        - \left( \bar{V}^{\varphi} {\vphantom{\big(}}^T \Delta^{\phi 0} \bar{V}^{\varphi} \right)_{ij} 
\end{align}
where
\begin{subequations}
\begin{align}
 A^\varphi &= U^\dagger U^\varphi + V^\dagger V^\varphi  \, , \\
 B^\varphi &= V^T       U^\varphi + U^T       V^\varphi  \, , \\
 R^\varphi &= \left( B^\varphi {A^\varphi}^{-1} \right)^* , \\
 \bar{U}^{\varphi} &= U^* + V R^\varphi , \\
 \bar{V}^{\varphi} &= V^* + U R^\varphi .
\end{align}
\end{subequations}

\subsection{Constraints}

It is often desirable to perform the energy minimization imposing a constraint on the expectation values of a selected 
set of operators. In this work, we choose to constrain the expectation values of the Bogoliubov quasiparticle state $\ket{\Phi}$ rather 
than those of the projected state $\ket{\Psi^{Z_0 N_0}}$.
There are many reasons motivating this choice.
First, it as the advantage of being cheaper from computational point of view as we only to evaluate the expectation values for a product state.
Second, it permits us to consider symmetry-breaking constraints, e.g.\ the constraint on the expectation value of a pair operator that breaks the gauge invariance. Finally, it is the Bogoliubov quasiparticle states, and not the projected states, 
that are considered as \emph{deformed} reference states in our calculations.
Nevertheless, this strategy has its drawbacks as different Bogoliubov states, obtained by using different constraints, may give the same projected 
state at the end of the VAPNP minimization. For example, an axially deformed Bogoliubov state may give a spherical particle-number projected state. 

First, let us specify that for a general operator $O$, we will constrain its hermitian average
\begin{equation}
 \tilde{O} = \frac12 \left( O + O^\dagger \right) ,
\end{equation}
to the desired value, let say $O_0$ 
\begin{equation}
 \elma{\Phi}{\tilde{O}}{\Phi} = O_0 .
\end{equation}

More generally, considering a set of constraints $\mathcal{C} \equiv \left\{ \tilde{O}_{k} , O_{k,0} \right\}$, the energy minimum in the hypersurface
defined by $\mathcal{C}$ can be obtained by using the modified gradient \cite{Egido80a,RS80a}
\begin{equation}
\label{eq:gradc}
 G = - \left[  \left( H^{20,Z_0 N_0}_{ij} - E^{20,Z_0 N_0}_{ij} \right) - \sum_{k}^{\mathcal{C}} \lambda_{O_k} \tilde{O}^{20}_k \right] , 
\end{equation}
where 
\begin{equation}
 \tilde{O}^{20}_{k,ij} = \frac{\elma{\Phi}{ \beta_j \beta_i  \tilde{O}_k}{\Phi}}{\scal{\Phi}{\Phi}} ,
\end{equation}
and the $\lambda_{O_k}$ are Lagrange multipliers that are obtained by solving self-consistently a system of linear equations 
(see Sec.\ref{sec:adjc}).

\subsubsection{Particle number}
\label{sec:zncons}
The most basic constraints we consider are the ones on the proton- and neutron-number operators. Because the operators are
hermitian, the constrained operators simply read
\begin{align}
 \tilde{Z} &= Z \, , \\
 \tilde{N} &= N  \, .
\end{align}
Let us remark that while VAPNP calculations do not require in principle such constraints due to the symmetry restoration embedded in the method, 
it is desirable to include them in the code for two reasons. 
First, they have to be used when doing plain HFB calculations. 
Second, such constraints usually improve the convergence of VAPNP calculations when using a sparse discretization of the gauge integrals and a large
step for the gradient descent. Otherwise, the gradient may not converge or change the number of particles to lower the energy.
While this particle-number constrained VAPNP is in principle not as general as the unrestricted one, it is in an interesting option whenever 
the computational time becomes problematic. A somewhat similar approach was used in \cite{Stoitsov07a}.
This is not a necessity though and we will consider in general full VAPNP calculations.

\subsubsection{Multipole deformations}
\label{sec:multi}
Together with pairing, deformation correlations are certainly the most prominent collective correlations exhibited by nuclei 
throughout the nuclear chart \cite{Bender06a}.
Therefore, to constrain the total deformation of the state $\ket{\Phi}$, we use the multipole operators 
\begin{equation}
 Q_{lm} = \sum_{ij} \elma{i}{r^l Y_{lm}}{j} c^\dagger_i c_j  \, ,
\end{equation}
where $Y_{lm}$ is a spherical harmonic. For now, we will consider the cases $l=2$ (quadrupole deformations), $l=3$ (octupole deformations)  
and $l=4$ (hexadecapole deformations).
Considering constraints on higher-order multipole operators would not present any theoretical difficulties. 
Their physical importance, however, is expected to be much smaller. 

Using the property
\begin{equation}
 Q_{lm}^\dagger = (-1)^m Q_{l-m} \, ,
\end{equation}
the hermitian average operators read 
\begin{equation}
 \tilde{Q}_{lm} = \frac12 \left( Q_{lm} + (-1)^m Q_{l-m} \right) \, ,
\end{equation}
and it sufficient to consider them for $m \ge 0$. 
Let us also remark in passing that the constraint on $\tilde{Q}_{21}$ has to be set to zero as it corresponds to the orientation of the state in space rather than a proper shape, 
as explained in \cite{RS80a}.

Instead of constraining the operators $\tilde{Q}_{lm}$ directly, we also consider the possibility to constrain the dimensionless operators \cite{Ryssens15a}
\begin{align}
 \tilde{\beta}_{lm} &= C_l \tilde{Q}_{lm} \, , \\
 C_l &=  \frac{4\pi}{3 R_0^l A}  \, ,
\end{align}
where $R_0 = 1.2 A^{1/3}$ fm. 

Also, when dealing with triaxial deformations, it is usually more intuitive to constrain the value of the triaxial parameters $(\beta,\gamma)$ rather than the average values of 
the multipole operators expressed in spherical coordinates. In that order, we can use the relations
\begin{align}
 \beta &= C_2 \sqrt{\langle \tilde{Q}_{20} \rangle^2 + 2 \langle \tilde{Q}_{22} \rangle^2 } , \\ 
 \gamma &= \arctan \left( \frac{\sqrt{2} \langle \tilde{Q}_{22} \rangle}{\langle \tilde{Q}_{20} \rangle} \right) , 
\end{align}
to determine the values of the constraints on $(\langle \tilde{Q}_{20} \rangle$, $\langle \tilde{Q}_{22} \rangle)$ necessary to obtain a state with the desired values 
of $(\beta,\gamma)$ in the end.

Finally, we want to stress that all the multipole deformations discussed here were defined using the ``bare'' multipole operators, i.e.\ using their textbook definitions.
While the use of effective operators adapted to the model space \cite{Suzuki95a}, or to the value of the flow parameter \cite{Hergert16a}, would be theoretically more appropriate, 
we are confident that the same variational space can be explored using the simpler bare operators.
In addition, it is important to note that these intrinsic deformations are not observable quantities. They are useful intermediate quantities that can be used to gain a
qualitative insight about the collective character of the nucleus and to generate quasiparticle vacua with different intrinsic configurations that can be eventually used in  
beyond-mean-field techniques, such as the projected Generator Coordinate Method \cite{RS80a}.

\subsubsection{Angular momentum}

The variation of Eq.~\eqref{eq:VAP} will in priority target the ground state energy. To better optimize the projected
energy with respect to the excited states, it is of advantage to use a constraint on the expectation value of 
the angular momentum components $J_k$, $k \in \left\{ x,y,z \right\}$. 
Such ``cranking'' constraints have been used for decades
in mean-field calculations to describe rotational bands of heavy nuclei \cite{RS80a,Afanasjev03a}. 
More recently, several MR EDF calculations \cite{Borrajo15a,Egido16a,Rodriguez20a} included cranked Bogoliubov states in their symmetry-projected configuration 
mixing and obtained a great improvement in the description of the low-lying excited states.

Given that the angular momentum component $J_k$ is hermitian, we obtain
\begin{equation}
 \tilde{J}_k = J_k \, . 
\end{equation}
Finally, let us remark that, as we consider only real matrices $U$ and $V$, we cannot constrain the real part of the expectation value of
the operator $J_y$ as it has only imaginary matrix elements within our choice of basis.

\subsubsection{Pair coupling}

While the VAPNP offers a better treatment of pairing correlations than HFB, it is still a method built upon a single
quasiparticle state which might not be sufficient whenever collective pairing fluctuations around the variational 
minimum are important.
In those cases, the description of the system can be improved by performing a configuration mixing of reference states
optimized with a pairing constraint \cite{Bally19a}.
In that order, we consider the constraint on an operator that couples pairs and which we generically label $[P^{\dagger}]$.
The hermitian average operator reads in that case
\begin{equation}
 \left[\tilde{P}^{\dagger}\right] = \frac12 \left( [P^{\dagger}] + [P] \right) \, .
\end{equation}
Several choices have been used in the literature concerning the precise type of coupling imposed. 
Frequently, the pairs are coupled in $JT$-scheme following the seniority model \cite{Talmi93a}
\begin{equation}
\left[\hat{P}_{\text{sen}}^{\dagger}\right]^{JT}_{M_{J}M_{T}} 
= \frac{1}{\sqrt{2}}\sum_{\breve{a}}\sqrt{2j_{a}+1}\left[c^{\dagger}_{\breve{a}}c^{\dagger}_{\breve{a}}\right]^{JT}_{M_{J}M_{T}} \, ,
\end{equation}
where the creation operators are $JT$-coupled according to 
\begin{align}
\left[c^{\dagger}_{\breve{a}}c^{\dagger}_{\breve{b}}\right]^{JT}_{M_{J}M_{T}} 
&= \frac{\sqrt{1-\delta_{\breve{a} \breve{b}}(-1)^{J+T}}}{1+\delta_{\breve{a}\breve{b}}}\sum_{\substack{m_{j_{a}}m_{j_{b}} \\ m_{t_{a}}m_{t_{b}}}} c^{\dagger}_{a}c^{\dagger}_{b} \\
& \times \langle j_{a}m_{j_{a}}j_{b}m_{j_{b}}|JM_{J}\rangle \langle \tfrac{1}{2}m_{t_{a}}\tfrac{1}{2}m_{t_{b}}|TM_{T}\rangle . \nonumber 
\end{align}
Both isoscalar ($T=0$, $J=1$) $pn$-pairing and isovector ($T=1$, $J=0$) $pp$-, $nn$-, and $pn$-pairing can be explored with these operators. 

Another possibility is to adopt  a more ``agnostic'' point of view and consider the coupling
\begin{equation}
\left[\hat{P}_{\text{agn}}^{\dagger}\right]^{JT}_{M_{J}M_{T}} 
= \sum_{\breve{a} \breve{b}} \left[c^{\dagger}_{\breve{a}}c^{\dagger}_{\breve{b}}\right]^{JT}_{M_{J}M_{T}} \, ,
\end{equation}
which can couple states coming from different orbits $\breve{a}$ and $\breve{b}$.

\subsubsection{Pairing field}

Another possibility to adjust the pairing content of the trial Bogoliubov vacuum is through the direct constraint of its pairing field, i.e.\ considering a constraint on the 
particle-number nonconserving one-body operator
\begin{equation}
\Delta_C = \sum_{ij} \Delta^{00}_{ij} c^\dagger_i c^\dagger_j \, ,
\end{equation}
where $\Delta^{00}_{ij}$ is the (state dependent) pairing field defined in Eq.~\eqref{eq:Dfield}.
In that case, the hermitian average operator reads
\begin{equation}
 \tilde{\Delta}_C  = \frac12 \left( \Delta_C + \Delta_C^\dagger \right)  \, .
\end{equation}

In practice, this constraint is equivalent to a rescaling of the pairing field, i.e.\ considering the modified field $\delta \Delta^{00}_{ij}$ with $\delta \in \mathbb{R}$,
when performing the iterative minimization \cite{Duguet20a}. This solution has the double advantage of possessing a straightforward implementation and offering 
a good numeric stability.

\subsection{Iterative procedure}
\label{sec:iter}

The minimization algorithm used is the combination of two iterative procedures. 
First, there are the global iterations, labeled with the letter $i$, that correspond to a new calculation of the particle-number projected 
energy gradient. Second, there are the local iterations within each iteration $i$, labeled with the letter $j$, that correspond to a new evolution of the wave function to enforce the constraints.
In the following, the iterations will be labeled $(i,j)$. The iterations start at $(0,0)$ and the last local iteration $j$ for a given $i$ will be noted $J$.

\subsubsection{Heavy-Ball method}
\label{sec:hball}
The Heavy-Ball algorithm (see Ref.~\cite{Ryssens19a} and references therein) is an efficient first-order minimization method that adds a momentum term to
the plain gradient in order to achieve a faster, and somewhat more reliable, convergence. 
More precisely, the gradient matrix used at iteration $(i,0)$ to evolve the wave function ($U^{(i-1,J)}$,$V^{(i-1,J)}$) is defined as 
\begin{equation}
 \bar{G}^{(i,0)} = \eta^{(i,0)} G^{(i,0)} + \mu^{(i,0)} \bar{G}^{(i-1,J)} , 
\end{equation}
with $\bar{G}^{(0,0)} = \bar{G}^{(0,J)} = 0$. The factors $\eta^{(i,0)}$ and $\mu^{(i,0)}$ can either be read from the input parameters and have fixed values during the iterative
procedure, i.e.\ $\eta^{(i,0)} = \eta^{(0,0)}$ and $\mu^{(i,0)} = \mu^{(0,0)}$, or be evaluated at each iteration using the formulae \cite{Ryssens19a} 
\begin{align}
 \eta^{(i,0)} &=  \left( \frac{2}{\sqrt{\varepsilon_\text{max}^{(i,0)}} + \sqrt{\varepsilon_\text{min}^{(i,0)}}} \right)^2 , \\
 \mu^{(i,0)} &= \left( \frac{\sqrt{\varepsilon_\text{max}^{(i,0)}} - \sqrt{\varepsilon_\text{min}^{(i,0)}}}{\sqrt{\varepsilon_\text{max}^{(i,0)}} 
                           + \sqrt{\varepsilon_\text{min}^{(i,0)}}} \right)^2 .
\end{align}
with $\varepsilon_\text{max}^{(i,0)}$ and $\varepsilon_\text{min}^{(i,0)}$ being approximations to the maximum and minimum eigenvalues
of the second derivative of the energy, respectively. The program includes two empirical recipes to build these approximations:
\begin{enumerate}
 \item Following the idea presented in Ref.~\cite{Robledo11a}, the eigenvalues of the HFB stability matrix are approximated using 
  the eigenvalues of $H^{11}$, i.e.\ the 11 part of the of Hamiltonian in the quasiparticle basis of the trial wave function \cite{RS80a}.
  More preciscely, we diagonalize $H^{11}$ at iteration $(i,0)$ and set
 \begin{align}
  \varepsilon_\text{max}^{(i,0)} &= 4 \max\left[ \text{diag}\left( H^{11,(i-1,J)} \right) \right]   , \\
  \varepsilon_\text{min}^{(i,0)} &= 2 \min\left[ \text{diag}\left( H^{11,(i-1,J)} \right) \right]   , 
 \end{align}
 where the factors were determined empirically. 
 \item We use the same principle but replacing the eigenvalues of $H^{11}$ by the ones of the single-particle Hamiltonian $h^{00}$, i.e.\
 \begin{align}
  \varepsilon_\text{max}^{(i,0)} &= 2 \max\left[ \text{diag}\left( h^{00,(i-1,J)} \right) \right]   , \\
  \varepsilon_\text{min}^{(i,0)} &= 2 \min\left[ \text{diag}\left( h^{00,(i-1,J)} \right) \right]   , 
 \end{align}
 where the factors were determined empirically. This method proved to be somewhat more stable when dealing with odd-even nuclei. 
\end{enumerate}

Finally, the iterative procedure stops when the gradient becomes smaller than a predetermined value $\epsilon_G$
\begin{equation}
  \frac{|| \bar{G}^{(i,0)} ||_F}{\eta^{(i,0)}} \le \epsilon_G ,
\end{equation}
or if the maximum number of global iterations is reached. 

\subsubsection{Adjusting the constraints}
\label{sec:adjc}

When imposing constraints during the energy minimization, the gradient has to be constructed according to Eq.~\eqref{eq:gradc}. 
The Lagrange multipliers $\lambda_{O_k}^{(i,j)}$ are calculated solving a system of linear equations
\begin{equation}
 A^{(i,j)} \lambda^{(i,j)} = b^{(i,j)} .
\end{equation}

First, assuming that the wave function $\ket{\Phi^{(i-1,J)}}$, which is associated with the matrices ($U^{(i-1,J)}$,$V^{(i-1,J)}$) in storage at the beginning
of iteration $(i,0)$, satisfies the set of of constraints $\mathcal{C}$, i.e.\
\begin{equation}
 \label{eq:cons}
 || \elma{\Phi^{(i-1,J)}}{\tilde{O}_k}{\Phi^{(i-1,J)}} - O_{k,0} || \le \epsilon_Q , \forall k , 
\end{equation}
with $\epsilon_Q$ being a numerical parameter, the system of equations is set with 
\begin{align}
  A_{kl}^{(i,0)} &= \text{Tr}\left( \tilde{O}^{20,(i-1,J)}_k \tilde{O}_l^{20,(i-1,J)}{\vphantom{\Big(}}^T \right) , \\
  b_k^{(i,0)} &=  \text{Tr}\left[ \left( H^{20,Z_0 N_0,(i-1,J)}_{ij} - E^{20,Z_0 N_0,(i-1,J)}_{ij} \right) \tilde{O}_k^{20,(i-1,J)}{\vphantom{\Big(}}^T \right] . 
\end{align}
With the Lagrange multipliers $\lambda_{O_k}^{(i,0)}$ thus obtained, the projected gradient $\tilde{G}^{(i,0)}$ is built according Eq.~\eqref{eq:gradc} and a first
evolution is performed to obtain the wave function $\ket{\Phi^{(i,0)}}$ associated with the matrices ($U^{(i,0)}$,$V^{(i,0)}$).

Then, if the evolved wave function $\ket{\Phi^{(i,0)}}$ does not satistify the condition \eqref{eq:cons}, we start a local iterative procedure setting
the system of linear equations with
\begin{align}
  A_{kl}^{(i,j)} &= \text{Tr}\left( \tilde{O}^{20,(i,j-1)}_k \tilde{O}_l^{20,(i,j-1)}{\vphantom{\Big(}}^T \right) , \\
  b_k^{(i,j)} &=  O_{k,0} - \elma{\Phi^{(i,j-1)}}{\tilde{O}_k}{\Phi^{(i,j-1)}}
\end{align}
and modifying the gradient as 
\begin{equation}
 \bar{G}^{(i,j)} = \bar{G}^{(i,j-1)} + \sum_{k}^{\mathcal{C}} \lambda_{O_k}^{(i,j)} \tilde{O}^{20,(i,j-1)}_k .
\end{equation}
After each calculation of the gradient $\bar{G}^{(i,j)}$, an evolved wave function $\ket{\Phi^{(i,j)}}$ is constructed. The iterative procedures stops whenever
$\ket{\Phi^{(i,j)}}$ satisfies the condition \eqref{eq:cons} or if the maximum number of local iterations is reached.

\subsubsection{Evolution of the wave functions}
\label{sec:evol}
The matrices ($U^{(i,j)}$,$V^{(i,j)}$) of the evolved quasiparticle state at the end of iteration $(i,j-1)$ are obtained as \cite{Egido95a}
\begin{align}
 U^{(i,j)} &= \left( U^{(i,j-1)} + {V^{(i,j-1)}}^{*} \bar{G}^{(i,j)}{\vphantom{\big(}}^* \right) {{L^{(i,j)}}^{-1}}^{\dagger} , \label{eq:evolU} \\
 V^{(i,j)} &= \left( V^{(i,j-1)} + {U^{(i,j-1)}}^{*} \bar{G}^{(i,j)}{\vphantom{\big(}}^* \right) {{L^{(i,j)}}^{-1}}^{\dagger} , \label{eq:evolV}  
\end{align}
where $L^{(i,j)}$ is the triangular matrix from the Cholesky factorization
\begin{equation}
L^{(i,j)} {L^{(i,j)}}^\dagger = 1_{d_\mathcal{M}} + \bar{G}^{(i,j)}{\vphantom{\big(}}^{T} {\bar{G}^{(i,j)}}{\vphantom{\big(}}^* ,
\end{equation}
and we set $(i,-1) \equiv (i-1,J)$.

\subsubsection{Iterative algorithm}
We give here a schematic overview of the iterative algorithm used in the code:
\begin{enumerate}[I)]
  \item Initialization of the global iterative procedure. 
   \begin{enumerate}[1)]
    \item $(U^{(0,J)},V^{(0,J)}) \equiv$ first trial wave function.
     If the maximum number of iteration $i_\text{max} > 0$, the initial wave function (read from file or randomly generated) has already been
     evolved to satisfy the constraints, with their matrix elements in the quasiparticle basis ${O^{20}}^{(0,J)}$ being stored.
    \item $\bar{G}^{(0,J)} = 0$.
   \end{enumerate}
  \item Loop over the global iteration number $i \in \llbracket 1, i_\text{max} \rrbracket$.
   \begin{enumerate}[1)]
    \item Computation of the particle-number projected gradient $\bar{G}^{(i,0)}$ using the wave function $(U^{(i-1,J)},V^{(i-1,J)})$,
          the projected gradient $\bar{G}^{(i-1,J)}$, and the matrix elements for the constraints ${\tilde{O}^{20}}{\vphantom{\Big(}}^{(i-1,J)}$.
   \item If $|| \bar{G}^{(i,0)} ||_F/\eta^{(i,0)} \le \epsilon_G$: go to III). 
   \item Evolution of the wave function taking into account the constraints.
    \begin{enumerate}[a)]
     \item Loop over the constraint iteration number $j \in \llbracket 0, j_\text{max} \rrbracket$
     \begin{enumerate}[i)]
      \item Computes $(U^{(i,j)},V^{(i,j)})$ and then the expectation values $\langle \tilde{O} \rangle^{(i,j)}$.
      \item If $|| \langle \tilde{O} \rangle^{(i,j)} - O_0 || \le \epsilon_O$: go to b).
      \item Computes the Lagrange multipliers $\lambda^{(i,j)}_O$ and with them updates $\bar{G}^{(i,j+1)}$.
     \end{enumerate}
     \item Storage of the final wave function $(U^{(i,J)},V^{(i,J)})$, projected gradient $\bar{G}^{(i,J)}$, and matrix elements of 
           the constraints ${\tilde{O}^{20}}{\vphantom{\Big(}}^{(i,J)}$, with $J$ being the last iteration of a).
    \end{enumerate}
   \end{enumerate}
  \item End of the iterative procedure
\end{enumerate}

%
%
\section{Examples}
\label{sec:examples}
%
%
\subsection{Unconstrained minima of $^{24,25}$Mg and $^{26}$Al with the USDB interaction}
\label{sec:exa1}

As first example, we perform unconstrained calculations for three nuclei in $sd$-shell with the USDB interaction \cite{Brown06a}: 
the even-even nucleus $^{24}$Mg, the odd-even nucleus $^{25}$Mg and finally the odd-odd nucleus $^{26}$Al. All three nuclei are located 
in the middle of the $sd$-shell and exhibit quadrupolar deformations in their ground state. In particular, $^{24}$Mg 
is found to have a triaxial minimum in valence space mean-field calculations \cite{Gao15a,Ryssens20a} as well as EDF calculations
that include angular momentum projection \cite{Bender08a,Rodriguez10a,Yao10a}. Similarly, $^{25}$Mg is found to be triaxial after the restoration of
the rotational invariance within the EDF framework \cite{Bally14a,Borrajo18a}. 
Unfortunately, advanced mean-field calculations of odd-odd mass nuclei are scarce
but axial calculations with the Gogny EDF found a prolate deformed minimum \cite{Hilaire07a}.

The oscillator frequency $\hbar \omega$ used in the calculations of each nucleus was determined using the formula\footnote{Note that the code uses this formula 
by default if no value of $\hbar \omega$ is entered in the Hamiltonian file (with \texttt{hamil\_type} = 1 or 2).
See the file \texttt{manual\_hamiltonian.pdf} for more information.}
 \cite{Suhonen07a}
\begin{equation}
\hbar \omega = \left( 45 A^{-1/3} - 25 A^{-2/3} \right), 
\end{equation}
where $A$ is the number of nucleons of the nucleus studied, and which gives the values $\hbar \omega = 12.60$ MeV ($b = 1.81$ fm) for $^{24}$Mg, 
$\hbar \omega = 12.47$ MeV ($b = 1.82$ fm) for $^{25}$Mg and $\hbar \omega = 12.34$ MeV ($b = 1.83$ fm) for $^{26}$Al.

We also specify that the triaxial deformations $(\beta,\gamma)$ were computed using the bare multipole operators, as explained in Sec.~\ref{sec:multi}.
In particular, this means that we did not use any effective operators or charges. 
This definition allows us to benchmark our results to the ones of Gao \etal~\cite{Gao15a} who made a similar choice.
On the other hand, it is important to remember that the values of $(\beta,\gamma)$ thus obtained will differ from the ones computed in
no-core calculations of the same nuclei. 
In this case, the difference does not come from the definition of the quadrupole operators but from the particles and orbits involved in their evaluation.

The calculations were carried out considering real general trial wave functions and the initial states were randomly generated.
For $^{25}$Mg, a quasiparticle was blocked on top of the random initial wave function.
For $^{26}$Al, we tried with and without the blocking of two quasiparticle excitations during the generation of the initial state. The two procedures
gave the same results.
It is important to remember that our general scheme allows the mixing of protons and neutrons such that
the wave function already includes odd-odd components even without an explicit blocking.
To bypass the problem of local minima, we repeated the calculations ten times before selecting the state giving the lowest energy for each nucleus.
In particular, we remark that in VAPNP calculations with wave functions that include proton-neutron pairing, we frequently found many local minima 
very close in energy.

In Table \ref{tab:miniALL}, we give the values of the total HFB energy $E_{\text{HFB}}$, the pairing part of the energy 
$E^\text{pair}_\text{HFB} = - \text{Tr}(\Delta \kappa^*)/2$,
and the triaxial deformation parameters ($\beta_{\text{HFB}}$,$\gamma_{\text{HFB}}$) for the unconstrained minima of the three nuclei. Also, for comparison, we give the value 
of the exact energy of the ground state $E_{\text{exact}}$ as obtained from the full diagonalization performed with the shell model code ANTOINE \cite{Caurier05a}.
As expected, all HFB energies are above the exact ones. More surprisingly, the HFB energies of $^{25}$Mg (odd-even) and
 $^{26}$Al (odd-odd) are in better agreement with the exact results than the HFB energy of $^{24}$Mg (even-even).
One problem with the $^{24}$Mg calculations is that the pairing collapses entirely as can be seen from its vanishing pairing energy.
All nuclei are found to be well deformed and to have a triaxial minimum.\footnote{Note here that the orientation of the nucleus is not fixed 
to correspond to a particular sextant of the ($\beta$,$\gamma$) plane. Therefore, the amount of triaxiality has to be determined relatively to the 
closest axial axis located  at $n\times60^\circ$, with $n$ an integer.} In particular, the minima of $^{25}$Mg and $^{26}$Al are about $22^\circ$ away from the 
closest axial axis.

Finally, we note that our results for $^{24}$Mg are in very good agreement with the ones of Gao \etal~\cite{Gao15a} also obtained with the USDB interaction.
In particular, using our definition for the deformation (see the caption of Table \ref{tab:miniALL})
 and considering the same number of digits, we obtain for the deformation $\beta=0.282$ and
$\gamma=11.98^\circ$ (relatively to the axis at $120^\circ$) which are very close to their values.
The energy is also well reproduced, differing only by 5 keV. Although one could expect an even better agreement between the two HFB calculations, we remark that
the value they give for the exact energy of $^{24}$Mg also differ from ours by 4 keV \cite{Gao15a}.
We do not know the origin of this discrepancy. It might comes from  a slight difference in the matrix elements
of the USDB interaction (we used the file provided on the webpage of the code ANTOINE).

\begin{table}
\begin{tabular}{cccc}
\hline
         &   $^{24}$Mg &  $^{25}$Mg & $^{26}$Al \\
\hline
 $E_\text{exact}$             &   -87.101 &  -94.399 & -105.749  \\[0.02cm]
 $E_\text{HFB}$               &   -80.960 &  -89.192 & -103.288  \\[0.02cm]
 $E^\text{Gao}_\text{HFB}$    &   -80.965 &          &           \\[0.02cm]
 $E_\text{VAPNP}$             &   -82.847 &  -90.732 & -102.711  \\[0.02cm]
 $E^\text{Gao}_\text{VAP-TA}$ &   -82.831 &          &           \\[0.02cm]
\hline
 $E^\text{pair}_\text{HFB}$   &    0.000  &  -0.582  & -2.675    \\[0.02cm]
 $E^\text{pair}_\text{VAPNP}$ &   -4.516  &  -3.680  & -3.006    \\[0.02cm]
\hline
 $\beta_\text{HFB}$               &   0.282   &  0.240   & 0.236     \\[0.02cm]
 $\beta^\text{Gao}_\text{HFB}$    &   0.281   &          &           \\[0.02cm]
 $\beta_\text{VAPNP}$             &   0.268   &  0.214   & 0.184     \\[0.02cm]
 $\beta^\text{Gao}_\text{VAP-TA}$ &   0.260   &          &           \\[0.02cm]
\hline
 $\gamma_\text{HFB}$               &   11.98   &   21.9   & 277.7     \\[0.02cm]
                                   &           &          & (37.7)    \\[0.02cm]
 $\gamma^\text{Gao}_\text{HFB}$    &   11.96   &          &           \\[0.02cm]
 $\gamma_\text{VAPNP}$             &   240.16  &  355.7   & 145.3     \\[0.02cm]
                                   &   (0.16)  &  (4.3)   & (25.3)    \\[0.02cm]
 $\gamma^\text{Gao}_\text{VAP-TA}$ &   0.09    &          &           \\[0.02cm]
\hline
\end{tabular}
\caption{Properties of the unconstrained HFB/VAPNP minima for the nuclei $^{24,25}$Mg and $^{26}$Al computed in the
$sd$-shell with the USDB interaction. 
The first row corresponds to the energies obtained by exact diagonalization using the shell model code ANTOINE.
The values of $\beta$ for Gao \etal~\cite{Gao15a} were computed using the relation
$Q = \sqrt{\frac{16\pi}{5 b^4}} \left( \frac{4\pi}{3 R_0^2 A} \right)^{-1} \beta$, where $R_0 = 1.2 A$ fm, $A$ is the total number of nucleons, 
 and $b = 1.81$ fm is the oscillator length. The values between parentheses represent the corresponding values of $\gamma$ in the first sextant of beta-gamma plane, 
i.e. with $ \gamma \in \left[ 0,60\right]$.
}
\label{tab:miniALL}
\end{table}%

Considering now the VAPNP calculations, the same properties, but now computed as expectation values for the projected state giving the lowest energy, are also given in Table \ref{tab:miniALL}.
The first observation is that the total energies of $^{24}$Mg and $^{25}$Mg are lowered compared to the HFB calculations 
and therefore get closer to the exact values. By constrast, the energy of $^{26}$Al is increased such that it gets worst at the VAPNP level.
This is a surprising result as VAP calculations are commonly thought as decreasing the energy compared to HFB calculations, 
which is indeed true in the vast majority of cases.
But there is no contradiction with the variational principle as only VAPNP calculations consider the variational space with the correct number of particles.
A particle-number projection analysis of the HFB minimum shows that the components $(Z=5,N=5)$ has an energy equal to $-101.147$ that is above the value obtained from the VAPNP.
Interestingly, the two other leading contributions\footnote{The components $(Z=5,N=5)$, $(6,6)$ and $(4,4)$ have weights approximately equal to $0.493$, $0.256$ and $0.244$, respectively.
The two other smaller components are $(3,3)$ and $(7,7)$.}
comes from the $N=Z$ even-even neighbors $(6,6)$ with an energy equal to $-131.697$ and $(4,4)$ with an energy equal to $-79.055$.

We note also that, in contrast with the HFB results, the pairing energy integrated over the gauge angles is non-zero
for all nuclei. It is a well-known fact that the methods based on the VAP for particle numbers provide a better description of the pairing correlations, in particular avoiding
its frequent collapse observed at the HFB level.

Concerning the deformation, we observe that the projected minima of the three nuclei are much less deformed than their HFB counterparts and in the case of 
$^{24}$Mg and $^{25}$Mg also become almost purely axial. 
Only $^{26}$Al conserves a large amount of triaxiality at the VAPNP level.
This reduction of triaxility for $^{24}$Mg was also observed by Gao \etal~ in their so-called VAP-TA calculations 
that include the isospin and particle-number projections in the variational process.
In Table \ref{tab:miniALL}, we compare their values to ours using our definition of the deformation parameters. It is remarkable that the 
two calculations agree so well between each other for the value of the projected energy and deformations in spite of the difference between the methods.
While their VAP calculations includes the full isospin projection in the variational minimization, our method  
consider more general trial wave functions. Indeed, we work here with general (real) quasiparticle states whereas their framework conserves
 the time-reversal invariance and the separation between protons and neutrons. In particular, we note that our VAPNP solution displays some proton-neutron mixing.

%
%
\subsection{Analysis of the Heavy-Ball algorithm for $^{24}$Mg}
\begin{figure}[t!]
\centering  
  \includegraphics[width=0.8\columnwidth]{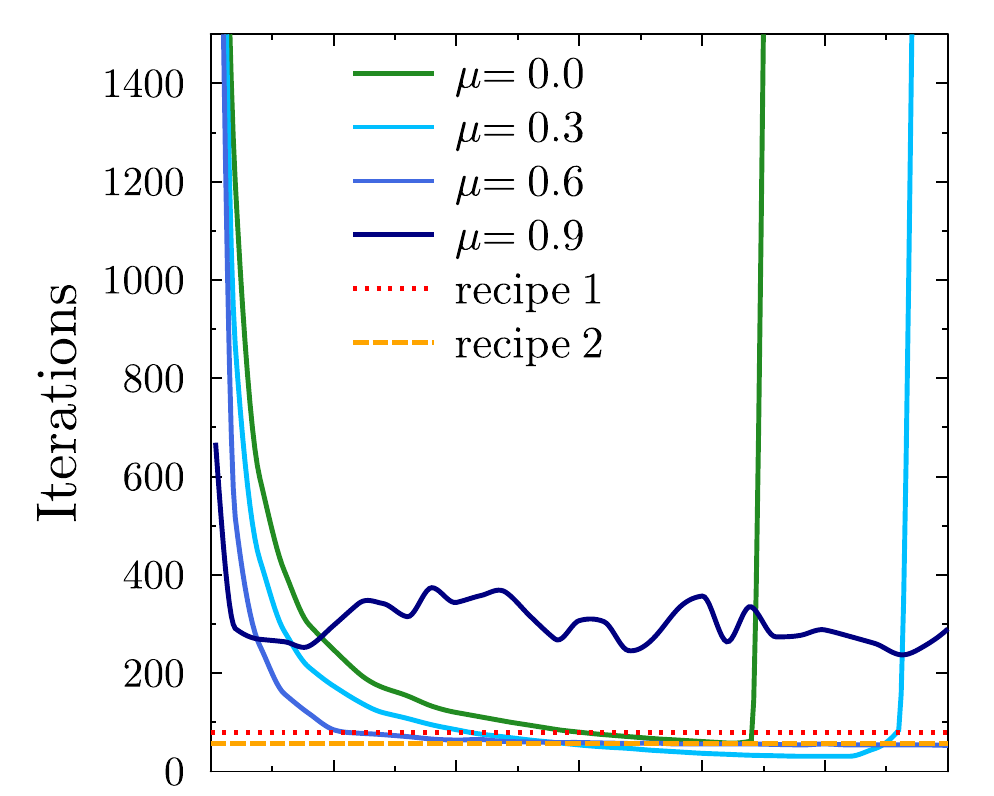} \\
  \includegraphics[width=0.8\columnwidth]{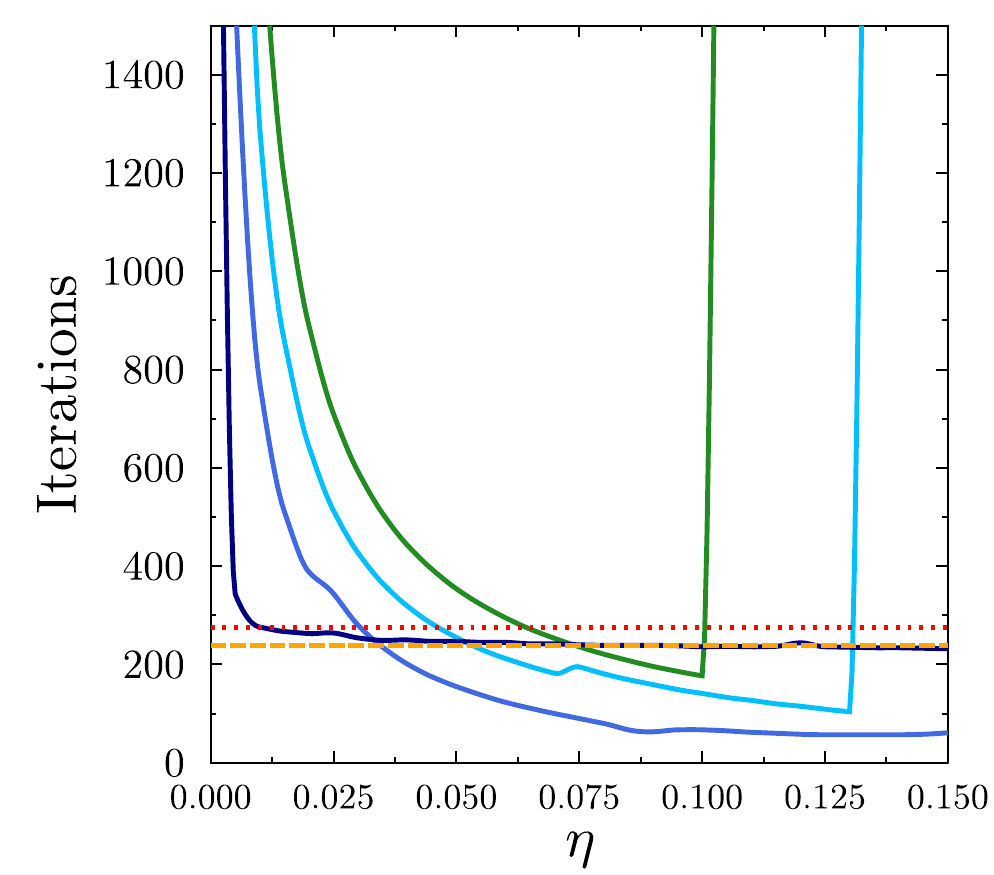} 
\caption{
\label{fig:iter}
(Color online)
Top: Evolution of the number of iterations required to converge a HFB calculation of the unconstrained minimum of $^{24}$Mg
as a function of the parameter $\eta$ for different gradient schemes. Bottom: Same for VAPNP calculations.
The convergence criterion is fixed at $\epsilon_G = 10^{-4}$.
The dashed lines represent the (optional) empirical recipes pre-implemented in the code.
}
\end{figure}

For our second example, we study the convergence of the Heavy-Ball (HB) algorithm for different choices for the parameters $\eta$ and $\mu$.
The calculations are performed using the same model space, interaction and oscillator parameters as the ones used in Sec.~\ref{sec:exa1}.
Taking the calculations of the HFB and VAPNP unconstrained minima of $^{24}$Mg as examples, we display in Fig.~\ref{fig:iter} 
the evolution of the number of iterations required to reach convergence as a function of the parameter $\eta$ for different fixed values of $\mu$.
The convergence criterion is fixed at $\epsilon_G = 10^{-4}$ in all calculations and all the calculations were started from the same initial wave function.
For comparison, we also show as dashed lines the results obtained for the two recipes discussed above that adapt $\eta$ and $\mu$ at 
each iteration. 

Looking first at the HFB results on the top panel, we observe that, as expected, increasing the value of $\eta$ decreases the number of iterations required
to reach convergence.
Nevertheless, it does so only up until a critical value $\eta_c$, e.g.\ $\eta_c \approx 0.11$ for the calculation with $\mu = 0$,
where the calculation does not converge anymore. When $\mu = 0.9$, however, the calculation always converges for the range of values of $\eta$ 
considered but does so in an erratic manner as a function of $\eta$.

Adding a momentum term, i.e.\ $\mu \ne 0$, has two main effects. First, for a given $\eta$, the calculations tend to converge faster. This is not a general rule 
however as the calculations with $\mu = 0.9$ are globally outperformed by the pure gradient descent with $\mu = 0$.
The second benefit of the presence of a momentum term is that it stabilizes the calculations by pushing up the value of $\eta_c$ 
over which the calculations do not converge anymore.

Comparing the HFB and VAPNP calculations, we see that the latter requires in general more iterations to converge than the former and also that the critical value $\eta_c$
is slightly smaller. Strangely, the calculation with $\mu = 0.9$ behaves better in the VAPNP calculation. Another remark, less apparent on the plot, is that the VAPNP calculations 
do not always converge to the exact same minimum. Indeed, there are two minima that are very similar (e.g.\ same deformation) but differ by 40 keV such that the calculations will transition
from one to the other. On the curves, the transitions can be seen as small steps at $\eta \approx 0.02$ for the calculation with $\mu=0.6$ and at $\eta \approx 0.07$ for the calculation with $\mu=0.3$.

Finally, we notice that the two empirical recipes introduced in Sec.~\ref{sec:hball} perform similarly in the two calculations. In the HFB case, they are extremely efficient and allow to 
reach convergence with less than 100 iterations. In the VAPNP case, however, they are not nearly as good and are outperformed by most of the other schemes.
The lesser performance of these recipes can be understood as they have been designed to use the quasiparticle/single-particle energies of the Bogoliubov state
at hand to optimize the unprojected gradient. Obviously, these quantities are not necessarily as well adapted to optimize the projected gradient built from the same state. 
Nevertheless, the two recipes still converge with an acceptable number of iterations without the need to manually fix the parameters.

%
%
\subsection{No-core calculations of $^{24}$Mg with a Chiral EFT interaction}

As final example calculation,\footnote{Contrarily to the two previous examples, the files used to perform this calculation are not provided in the repository.} we consider again the nucleus $^{24}$Mg but this time performing a no-core calculation in a model space characterized by $e_{\text{max}} = (2 n + l)_{\text{max}} = 8$, i.e. containing 9 major oscillator shells, 
and $\hbar \omega = 20$ MeV. The chiral Hamiltonian used contains a 2N interaction at next-to-next-to-next-to leading order (N3LO) \cite{Entem03a}, a 3N interaction at N2LO \cite{Navratil07a}, and was evolved through Similarity 
Regularization Group (SRG) techniques \cite{Roth12a}. In addition, the 3N part of the Hamiltonian was recasted following the procedure explained in \cite{Frosini20a} using as reference state 
a spherical HFB state constructed for $^{24}$Mg.
In this special case, this procedure is similar in spirit to the well-known NO2B \cite{Roth12a}.

\begin{figure}[t!]
\centering  
  \includegraphics[width=0.8\columnwidth]{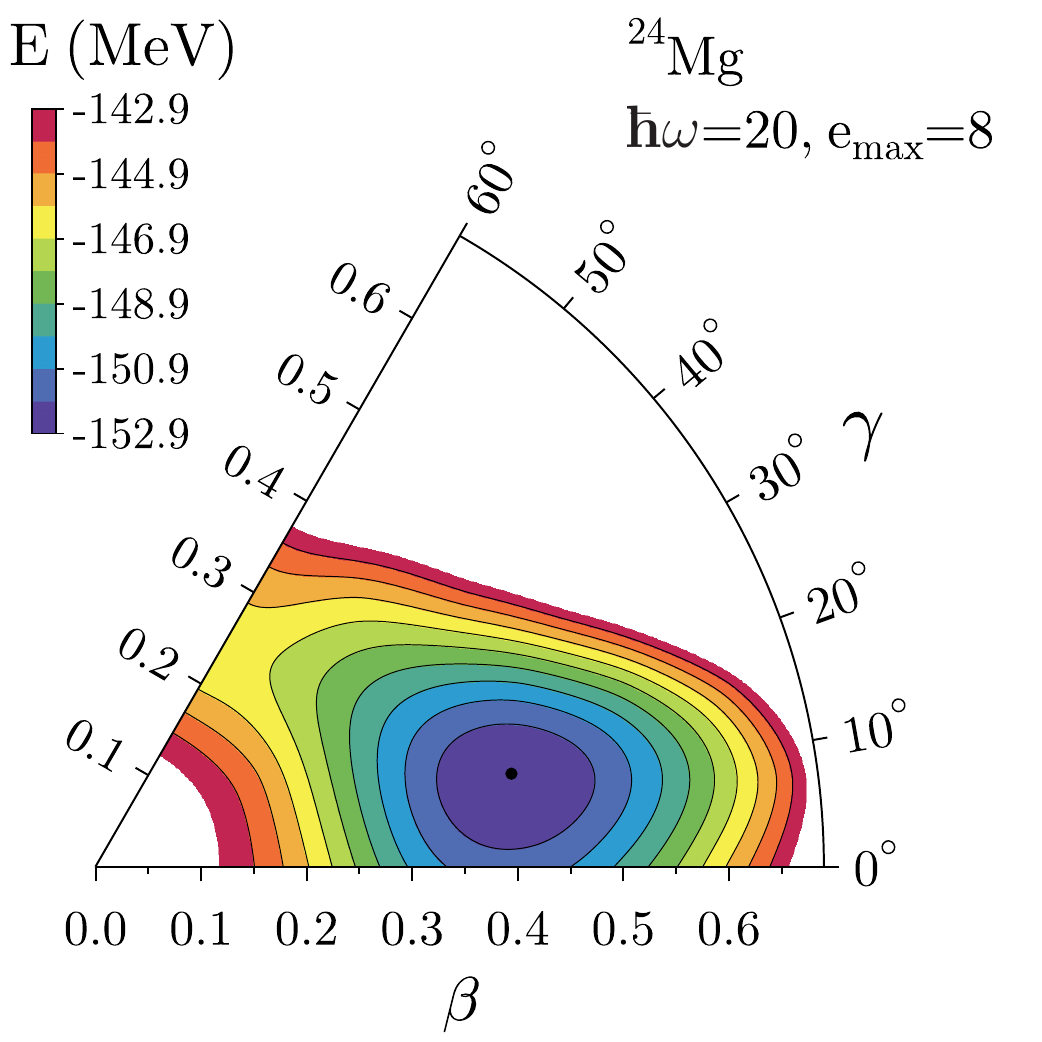} 
\caption{
\label{fig:abini}
(Color online)
Energy surface in the first sextant of the ($\beta$,$\gamma$) plane for the HFB calculations performed in a model space with $e_{\text{max}} = 8$, $\hbar \omega = 20$ and using the chiral Hamiltonian
described in the text. The black lines are separated by 1 MeV, starting from the minimum represented by a black dot.
}
\end{figure}

In figure \ref{fig:abini}, we display the HFB energy surface in the first sextant of the triaxial ($\beta$,$\gamma$) plane.
All calculations were started from randomly generated wave functions breaking all symmetries but the parity\footnote{When starting from parity breaking states, most of the calculations converge towards
positive parity wave functions at the end of the minimization, such that the energy surface is mostly unchanged. However, a few final wave functions break the parity 
and would require a subsequent parity projection that is outside the scope of the present article.}
and were constrained to have $\langle \tilde{O}_{21} \rangle = 0$. 
As can be seen, the HFB minimum is triaxial with a deformation  ($\beta=0.404$, $\gamma=12.62^\circ$) and an energy $E_{\text{HFB}} = -152.937$ MeV. Interestingly, the value for
$\gamma$ is in very good agreement with the one obtained in the valence-space USDB calculations discussed previously. 
On the other hand, the value for $\beta$ is slightly larger, which is not totally surprising as we consider now all the nucleons as being active in a much larger model space.
In other words, a fair comparison between the values of $(\beta,\gamma)$ obtained in the valence-space and no-core calculations would have required the use of proper effective quadrupole operators 
in the valence-space calculations.
 We also recall that state-of-the-art EDF calculations only obtain an axial minimum
at the HFB or VAPNP level \cite{Bender08a,Rodriguez10a,Yao10a}. Here, a VAPNP calculation for the minimum gives a very similar result for the deformation ($\beta=0.403$, $\gamma=12.45^\circ$), but
with an energy $E_{\text{VAPNP}} = -157.010$ MeV. Therefore, we do not observe any reduction of the triaxiality, contrary to the valence-space calculations. 
We remark in passing that while the pairing collapses in the wave function of the HFB minimum, the pairing energy (integrated over the gauge angles) of the VAPNP solution
is completely dominated by the neutron-proton contribution even if the chiral Hamiltonian breaks explicitly the isospin.

The energy of the HFB and VAPNP minima are more than 40 MeV above the experimental ground-state energy, $E_{\text{exp}} = -198.257$ MeV \cite{Wang17a}.
This is not surprising because these methods miss many important correlations as shown, for example, in many-body perturbation theory calculations performed on top of HFB minima \cite{Tichai18a}. 
This deficiency could be corrected by unitarily transforming the Hamiltonian to include the necessary correlations, as done in Ref.~\cite{Yao20a}
 using the In-Medium Similarity Renormalization Group method \cite{Hergert16a}. 
Nevertheless, it is interesting to remark that the energy surface is very similar to what can be obtained in EDF calculations at the multi-reference level \cite{Bender08a,Borrajo18a}.
It might be that even if absolute quantities (e.g.\ the total energy) are not well reproduced at the HFB level, relative quantities (e.g.\ the energy difference between deformed states and thus
the deformation adopted by the minimum) are.

Finally, we remark that an $e_{\text{max}} = 10$ calculation gives a minimum with a deformation ($\beta=0.407$, $\gamma=12.65^\circ$) and an energy $E_{\text{HFB}} = -153.179$ MeV, thus the 
calculation is relatively well converged as a function of the number of shells (for a fixed value of $\hbar \omega = 20$ MeV).

%
%
\section{Conclusions}
\label{sec:conclu}

In this article, we presented the numerical code \TAURUSvap~that solves the variation after particle-number projection equations
for real general Bogoliubov quasiparticle states represented in a spherical harmonic oscillator basis.
We explained the theoretical framework within which the code operates and outlined its principal features. Additional information
can also be found on the GitHub repository associated with this article. 

We hope that making this advanced variational solver publicly available will be of interest to the nuclear structure community in general.
In particular, \TAURUSvap~can be used to carry out constrained variational calculations in both valence-space and no-core model spaces with
general Hamiltonians containing up to two-body terms. Moreover, the \textsf{MPI} implementation included in the code allows one to use 
parallel supercomputers to run large model space calculations. 

While the numerical code \TAURUSvap~already allows the realization of advanced variational calculations, many developments are possible in the future 
such as the inclusion of additionial symmetry projections in the variational scheme or the use of the full three-body interaction. 
Nevertheless, the next logical step will be the publication of the companion code \TAURUSpav~that, considering the same theoretical framework,
 performs various symmetry projections using the output wave functions produced by \TAURUSvap.

%
%
\section{Using the code}
\label{sec:struct}

\subsection{Source files and compilation}
The numerical code \TAURUSvap~is divided into different files containing the main program, the various modules, and specific mathematical routines.
The list all the files, in the order of which they should be compiled, is the following: \\[0.2cm]
$\tt module\_constants.f90       $ \\
$\tt module\_mathmethods.f90     $ \\ 
$\tt module\_parallelization.f90 \text{ (only if \textsf{MPI})}$ \\
$\tt module\_nucleus.f90         $ \\
$\tt module\_basis.f90           $ \\
$\tt module\_hamiltonian.f90     $ \\
$\tt module\_wavefunctions.f90   $ \\
$\tt module\_fields.f90          $ \\
$\tt module\_particlenumber.f90  $ \\
$\tt module\_pairs.f90           $ \\
$\tt module\_angularmomentum.f90 $ \\
$\tt module\_multipoles.f90      $ \\
$\tt module\_radius.f90          $ \\
$\tt module\_operators.f90       $ \\
$\tt module\_projection.f90      $ \\
$\tt module\_constraints.f90     $ \\
$\tt module\_gradient.f90        $ \\
$\tt module\_initialization.f90  $ \\
$\tt subroutines\_pfaffian.f     $ \\
$\tt taurus\_vap.f90             $ \\[0.2cm]

To use the MPI implementation, it is first required, before compilation, to remove the comment flags ``\texttt{!cmpi}'' present in the Fortran files. 
This can be done easily using the command\footnote{On BSD-based systems, it is necessary to use instead $\texttt{sed -i "" "s/\textbackslash !cmpi//g" src/*f90}$}
\begin{equation*}
\texttt{sed -i "s/\textbackslash !cmpi//g" src/*f90}
\end{equation*}
while being in the main directory of the respository.\footnote{As this command changes the source files in an irreversible manner, we recommend to first create
a temporary copy of the source files before using it. Otherwise, it will be necessary to discard the changes in the local git repository before the next compilation of the code
in sequential mode (i.e.\ without MPI).}

On the GitHub repository, we provide a makefile to compile the code that takes in entry two arguments: \texttt{FC} to specify the compiler
and \texttt{TH} to specify if we want to include OpenMP threading or not. If no argument is entered, the script will use some default values.
The correct values for the argument \texttt{FC} are: \texttt{gfortran} (default), \texttt{ifort}, \texttt{mpiifort} and \texttt{mpif90}. 
On the other hand, the argument \texttt{TH} can take the values \texttt{omp} or \texttt{none} (default). 
For example, if one wants to compile \TAURUSvap~using the \texttt{gfortran} compiler and without OpenMp, one has to execute the command
\begin{equation*}
\texttt{make FC=gfortran TH=none}
\end{equation*}

For completeness, we also provide in the repository a bash script that offers comparable compilation capabilities. 
More information about how to use this script or the makefile can be found in the file \texttt{README.md} present in the main directory
of the repository.

\subsection{Execution}
Once in the directory containing the executable file, the code can be run by typing the command
\begin{equation*}
\texttt{./taurus\_vap.exe < input.txt}
\end{equation*}
where \texttt{input.txt} is the STDIN file containing the input parameters. The details concerning the format of STDIN can be found in the 
file \texttt{manual\_input.pdf} present in the directory \texttt{extras}. The code also requires, in the same directory, the file defining the Hamiltonian (and model space), 
the name of which is written in the STDIN, and the various files containing its matrix elements. 
The details concerning the format of Hamiltonian files can be found in the file \texttt{manual\_hamiltonian.pdf} present in the directory \texttt{extras}.

To simplify the execution of the code, we provide the script \texttt{launch.sh} that performs all the necessary steps to run a calculation. 
To use it, go to the main directory of your copy of the repository and type the command 
\begin{equation*}
\texttt{bash launch.sh}
\end{equation*}

During its execution, the code prints various information (e.g. inputs and model space used, expectation value of the energy at each iteration, etc.) in the STDOUT. 
We recommend to store the printing in a file, for example \texttt{output.txt}, by typing 
\begin{equation*}
\texttt{./taurus\_vap.exe < input.txt > output.txt}
\end{equation*}
or
\begin{equation*}
\texttt{bash launch.sh > ouput.txt}
\end{equation*}

Additionally, the code will produce other files containing relevant information such as the occupation numbers or the eigenvalues 
of the single-particle Hamiltonian. More importantly, the code will write the final wave function obtained at the end of the iterative procedure in a file.
The names of all the files produced during a run are recalled in the STDOUT.
See the file \texttt{extras/manual\_input.pdf} for more details.

\subsection{Time and memory requirements}
The runtime of the code will depend on three main factors:
\begin{enumerate}
 \item The number of non-zero two-body matrix elements of the Hamiltonian stored, which is related to the size of the model space considered. 
 \item The number of discretization points used in the particle-number projection. 
 \item The number of iterations required to converge the calculation, which will depend on the initial wave function and the parameters used for the
       calculation (e.g.\ the criterion for the convergence of the gradient).
\end{enumerate}
For that reason, it is difficult to give a precise estimate for the time required to complete a run.
Nevertheless, from experience we can say that a run in a valence space will take between a few tenths of a second and a few minutes. 
Similarly, no-core calculations  including only a few shells, e.g.\ $e_\text{max} = 8$, will be take only a few minutes.
Computations in larger model space, however, may take several hours or even days to complete. 

\begin{figure}[t!]
\centering  
  \includegraphics[width=\columnwidth]{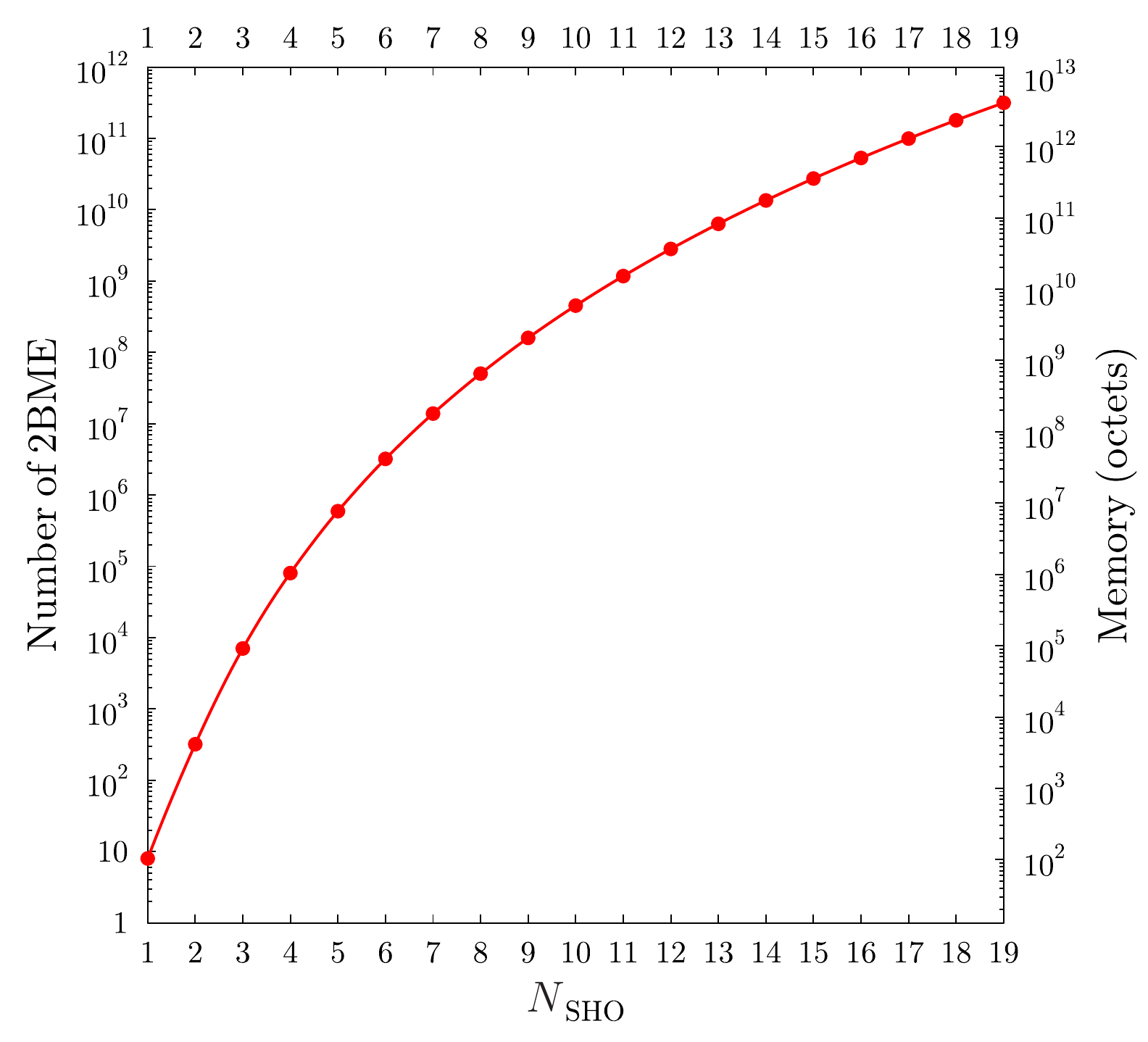}
\caption{
\label{fig:tbme}
(Color online)
Evolution of the number of two-body matrix elements that have to be stored, using the strategy explained in Sec.~\ref{sec:2BME}, 
as a function of the number of oscillator shells $N_{\text{SHO}}$ included in a no-core calculation.
All matrix elements not forbidden for symmetry reasons are assumed to be non-zero.
}
\end{figure}

On the other hand, the amount of memory required for a calculation is almost entirely driven by the number of non-zero two-body matrix elements of the Hamiltonian stored. 
In Fig.~\ref{fig:tbme}, we give the maximum amount of memory required to store the two-body matrix elements (2BME) of $H$ in single-precision as a function of 
the number of oscillator shells $N_{\text{SHO}}$ included in a no-core calculations.
The results were obtained using the storage strategy explained in Sec.~\ref{sec:2BME} assuming that all matrix elements not forbidden for symmetry
reasons are non-zero. As can be seen, the memory required grows rapidly with the number of shells but remains manageable even for large values 
of $N_{\text{SHO}}$ especially considering the parallelization capabilities of our modern computers (see Sec.~\ref{sec:para}).

As a conclusion, we will remark that a run in a small valence space has, for all intents and purposes, no system requirements
and can be completed in a short amount of time even on a laptop.
By contrast, no-core calculations in large model spaces are computationally heavy and will require
the use of a parallel supercomputer or computing cluster.

\subsection{Parallelization}
\label{sec:para}
In order to make the large model space calculations possible, the code integrates three layers of parallelization, based on \textsf{MPI} and \textsf{OpenMP}, that can be used
separetely or together. 
The parallelization scheme depends on three parameters that are handed to the code: the total number of \textsf{MPI} proccesses $N_\text{worldsize}$, 
the desired number of \textsf{MPI} processes in a team $N_\text{teamsize}$, and the number of \textsf{OpenMP} threads per \textsf{MPI} proccess $N_\text{threads}$.
Then, the calculations are parallelized following the strategy:
\begin{enumerate}
  \item The intregral over gauge angles of the PNP is parallelized by distributing a different set of angles to calculate to each of the $N_\text{teams}$ teams. 
  After the calculations of all the angles, the results are reduced towards the global master process to obtain the final projected quantities and continue the iterative procedure.
  \item Within each team, each process stores only a subset of the 2BME and uses them to compute the fields $\Gamma^{0\varphi}$, $\Delta^{0\varphi}$ and $\Delta^{\varphi 0 *}$.
  After the calculations, the results are reduced towards the team master process to obtain the final matrix elements of the fields at angle $\varphi$.
  \item Within each process, the sum over the stored 2BME in the calculations of the fields is distributed among the threads. 
  After the calculations, the results are reduced towards the master thread to obtain the matrix elements of the fields for the process at angle $\varphi$.
\end{enumerate}

\subsubsection{Distribution of the processes}
To distribute the processes among the teams, the code computes the euclidean division 
\begin{equation}
 N_\text{worldsize} = p N_\text{teamsize} + r  \quad (0 \le r < N_\text{teamsize})  
\end{equation}
and sets 
\begin{equation}
  N_\text{teams} = p + 1 - \delta_{r0} .
\end{equation}
Then, the number of processes assigned to each team is defined as
\begin{equation}
N_\text{myteamsize} = \left\{ 
\begin{array}{ll}
 N_\text{teamsize}  & \text{ [first $(N_\text{teams} - 1)$ teams]} \\
 r + \delta_{r0} N_\text{teamsize} &  \text{ [last team]}
\end{array} \right.
\end{equation}
The sets of $N_\text{teamsize}$ processes are assigned sequentially according to their global rank.

\subsubsection{Distribution of the angles}
To distribute the $N_\text{angles} = M_{\varphi_N} M_{\varphi_Z}$ (see Sec.~\ref{sec:discPNP}) angles among the teams, the code computes the euclidean division
\begin{equation}
 N_\text{angles} = p N_\text{teams} + r  \quad  (0 \le r < N_\text{teams}) .
\end{equation}
Then, the number of angles attibuted to each team is defined as 
\begin{equation}
N_\text{myangles} = \left\{ 
\begin{array}{ll}
 p + 1   & \text{ [first $r$ teams]} \\
 p       & \text{ [last $(N_\text{teams}-r)$ teams]}
\end{array} \right.
\end{equation}
The sets of $N_\text{myangles}$ angles are assigned sequentially according to their order in the loop over gauge angles. 

\subsubsection{Distribution of the matrix elements}
To distribute the $N_\text{2BME}$ two-body matrix elements of $H$ among the members of a team, the code computes the euclidean division
\begin{equation}
 N_\text{2BME} = p N_\text{myteamsize} + r  \quad  (0 \le r < N_\text{myteamsize}) .
\end{equation}
Then, the number of matrix element assigned to each member is defined as 
\begin{equation}
N_\text{my2BME} = \left\{ 
\begin{array}{ll}
 p + 1   & \text{ [first $r$ members]} \\
 p       & \text{ [last $(N_\text{myteamsize}-r)$ members]}
\end{array} \right.
\end{equation}
The sets of $N_\text{my2BME}$ matrix elements are assigned sequentially according to their order of reading in the Hamiltonian input file. 

\subsection{Dependencies}

The code requires the \textsf{BLAS} and \textsf{LAPACK} libraries.
In addition, it is recommanded to use a recent compiler as the code includes a few Fortran 2003/2008 commands
that might not be implemented in compilers that are too old.

%
%
\section{Additional technical details}
\label{sec:detail}

\subsection{Seed generation}
The wave function that is used as initial guess in the iterative procedure can either be read from a file or generated randomly, depending on the
value of the input parameter \texttt{seed\_type} (see the manual). If the user chooses to start the calculation from a random initial state, the following strategy
is used for its generation:
\begin{enumerate} 
 \item The seed of the pseudorandom generator is initialized by executing "\texttt{call random\_seed()}", which generates a seed based
  on the system's time.
 \item The occupations $v^2$ in the SHO basis, assumed for now to be the canonical basis, 
 are generated randomly and from them the matrices $U_c$ and $V_c$ of the Bogoliubov transformation.
 \begin{enumerate}[i)]
  \item In the case of a BCS or general state, the occupations correspond to a fully paired vacuum ($0 \le v^2 < 1$) and the single-particle states 
  are paired with their time-reversal partners ($a$ with $-a$). In addition, if the state is a spherical BCS wave function,
  there is an equal filling of all single-particle states belonging to the same orbit.
  At the end of the process, the occupations are scaled such that the states have the correct 
  numbers of particles $Z$ and $N$ on average.
  \item In the case of a Slater determinant, $Z$/$N$ randomly selected proton/neutron single-particle states are fully occupied ($v^2 = 1$).
  \end{enumerate}
 \item A random orthogonal transformation $D$ is generated by diagonalizing a randomly generated real symmetry matrix $S$ and forming the matrix made of its
 eigenvectors.
  \begin{enumerate}[i)]
   \item In the case of a BCS state (spherical or not), $D$ is assumed to be the identity matrix $1_{d_{\mathcal{M}}}$.
   \item In the case of a general state, the symmetric matrix $S$ has the dimensions $d_{\mathcal{M}} \times d_{\mathcal{M}}$.   
   \item If some symmetries are to be conserved (such as parity or the separation between protons and neutrons), several smaller matrices $S_i$
    with the dimensions of the symmetry blocks are generated and diagonalized. Ultimately, $D$ is built from the combinations of their eigenvectors.
  \end{enumerate}
 \item A random orthogonal transformation $C$ is generated following the same principles. In the case of a Slater determinant, $C$ is assumed to be the identity matrix $1_{d_{\mathcal{M}}}$.
 \item The final Bogoliubov matrices of the inital state in the SHO basis are generated, using the Bloch-Messiah-Zumino theorem \cite{RS80a}, as $U = D U_c C$ and $V = D^* V_c C$.\footnote{Note that
   for a state with proton-neutron mixing, only the average number of nucleons $A$ will be conserved after applying the transformations $D$ and $C$ defined above.}
\end{enumerate}

\subsection{Discretization of the particle-number projection operator}
\label{sec:discPNP}
The integral over gauge angles appearing in the particle-number projection operator is discretized following the idea of Fomenko \cite{Fomenko70a}.
More precisely, taking the neutrons as example (the exact same discretization is used for protons), we use the $M_{\varphi_N}$-point quadrature
\begin{equation}
 \label{eq:flamenko}
  \mathcal{P}^{N_0}_{M_{\varphi_N}}
= \frac{1}{{M_{\varphi}}} \sum_{m=1}^{{M_{\varphi_N}}} 
  e^{\iunit  \alpha \pi f(m,M_{\varphi_N}) \, (N - N_0)} ,
\end{equation}
where
\begin{equation}
 \alpha  = \left\{
 \begin{array}{cl}
  1 &\text{ if no p-n mixing,} \\[0.2cm]
  2 &\text{ otherwise,} 
 \end{array}
 \right. 
\end{equation}
\begin{equation}
 f(m,M_{\varphi_N}) = \left\{ 
 \begin{array}{cl}
  \frac{m-1}{M_{\varphi_N}} &\text{ if $M_{\varphi_N}$ is odd,} \\[0.2cm]
  \frac{m-1/2}{M_{\varphi_N}} &\text{ if $M_{\varphi_N}$ is even.} 
 \end{array}
 \right.
\end{equation}
The selection of a different discretization for the angles as a function of the parity of $M_{\varphi_N}$ is done
to avoid the numerical evaluation of the rotated quantities at angle $\frac{\pi}{2}$ whatever the value of $M_{\varphi_N}$. 
Indeed, when going through the angle $\frac{\pi}{2}$, one might have to evaluate a fraction where both the numerator and
denominator can become arbitrarily close to zero, which is numerically dangerous \cite{Anguiano01a}.

More details on the mathematical aspects of the discretized operator $\mathcal{P}^{N_0}_{M_{\varphi_N}}$ as well as an 
analysis of its convergence behavior can be found in Ref.~\cite{Bally20a}.

\subsection{Evaluation of overlaps}
\label{sec:pfaf}
The projection onto good particle numbers $Z$ and $N$ requires the evaluation of the overlap $\scal{\Phi}{\Phi(\varphi_Z,\varphi_N)}$ between
the state $\ket{\Phi}$ and the gauge rotated state $\ket{\Phi(\varphi_Z,\varphi_N)}$ for all the discretized values of $(\varphi_Z,\varphi_N)$.
The evaluation of overlaps between general quasiparticle states can be performed using the Pfaffian algebra \cite{Robledo09a}. In this work, we
use the particular formula given in Eq.~(60) in Ref.~\cite{Avez12a} that presents several advantages, the first of which being that fully occupied/empty single-particle 
states, once identified, can be treated rather straightforwardly.\footnote{We remark that Slater determinants, which represent the special case where all the single-particle 
states are either fully occupied or fully empty, can also be treated using this formula.}

The most direct way to identify those single-particle states is by constructing the canonical basis of $\ket{\Phi}$. To do so, we first
diagonalize the one-body density $\rho$ and transform the one-body pairing tensor $\kappa$ in this basis. This change of basis is sufficient
to transform all the blocks of $\kappa$ corresponding to non-degenerate eigenvalues of $\rho$ into their canonical form. In case there are remaining 
blocks of $\kappa$ that are not yet in their canonical form, 
we proceed to a Schur decomposition of each of these blocks, which automatically put them in their
canonical form given $\kappa$ is a real antisymmetric matrix. In the end, both $\rho$ and $\kappa$ are transformed in this new basis that both
diagonalize $\rho$ and put $\kappa$ in its canonical form.

Once the canonical basis has been constructed, it is easy to identify the fully occupied/empty states of $\ket{\Phi}$ in this basis by searching
for the states $i$ that have an occupation $v^2_i$, i.e.\ an eigenvalue of $\rho$, above/below a desired numerical value. In the code,
a state $i$ is considered fully occupied if $v^2_i \ge 1 - \epsilon$, with $\epsilon$ being a numerical parameter set by default to $10^{-8}$ but that
can also being chosen via the parameter \texttt{seed\_occeps} in the input file. 
On the other hand, a state $i$ is considered fully empty if $v^2_i \le \epsilon$.

After identification, the fully occupied/empty single-particle states in the canonical basis are removed from the calculations of the Pfaffian following the 
methodology presented in Ref.~\cite{Avez12a}.
The big advantage of removing the empty states is that it reduces the dimensions of the matrices and therefore reduces the computational time required to 
compute the Pfaffian.  Nevertheless, in some rare occasions, it becomes necessary to include the empty states to avoid having an overlap matrix of the 
single-particle states that is non-invertible.

On the numerical side, the Pfaffian is computed using the efficient routines provided in the library \textsf{PFAPACK} \cite{Wimmer12a}. 
We integrated the routines required for our calculations in the file \texttt{subroutines\_pfaffian.f}.

\subsection{Two-body matrix elements of the Hamiltonian}
\label{sec:2BME}

The number of two-body matrix elements of the Hamiltonian in the SHO single-particle basis can be very large when the model space considered
is made of many oscillator shells. Unfortunately, a large number of matrix elements poses several difficulties. 
First, the amount of memory required to store the matrix elements is directly proportional to the number of matrix elements and can rapidly represent hundreds of gigaoctets
or more. Second, in no-core calculations, the run-time within a single iteration is entirely dominated by the calculations of the fields 
($\Gamma^{0\varphi}$, $\Delta^{0\varphi}$ and $\Delta^{\varphi 0 *}$) and is computationally intensive.
Finally, the decoupling of the $J$-scheme matrix elements read in the input files to the $m$-scheme matrix elements used in the calculation,\footnote{When 
performing valence-space calculations, the matrix elements are read in $JT$-scheme and decoupled using the adapted expression \cite{Suhonen07a}.}
i.e.\ \cite{Suhonen07a}
\begin{equation}
\begin{split}
h^{(2)}_{abcd} &\equiv \elma{ab}{h^{(2)}}{cd} \\
               &= \sum_{J M_J} (j_a m_{j_a} j_b m_{j_b} | J M_J ) (j_c m_{j_c} j_d m_{j_d} | J M_J ) \\
               &\phantom{=} \times \left[ \mathcal{N}_{\hat{a}\hat{b}}(J) \mathcal{N}_{\hat{c}\hat{d}}(J) \right]^{-1}  \elma{\hat{a}\hat{b},J}{h^{(2)}}{\hat{c}\hat{d},J}
\end{split}
\end{equation}
where $\elma{\hat{a}\hat{b},J}{h^{(2)}}{\hat{c}\hat{d},J}$ is a $J$-scheme matrix element
and 
\begin{equation}
 \mathcal{N}_{\hat{a}\hat{b}}(J) = \frac{\sqrt{1+\delta_{\hat{a}\hat{b}} (-1)^J}}{1 + \delta_{\hat{a}\hat{b}}}
\end{equation}
is a normalization factor, can also become prohibitively long. Similar expression is used when decoupling a $JT$-scheme matrix elements

To overcome these problems, the main strategy is to store and use in the calculations the minimal number of matrix elements possible. The first obvious step is to store only
the non-zero matrix elements $\bar{h}_{abcd}$ and keep track of the corresponding indices $a,b,c,d$. 
In that order, for each $m$-scheme matrix element $\bar{h}_{abcd}$ above a certain numerical cutoff (by default $10^{-16}$), we store:
\begin{enumerate}
 \item The matrix element $\bar{h}_{abcd}$ in a one-dimensional 4-byte real array.
 \item The quartet of indices ($a,b,c,d$) in a one-dimensional 2-byte integer array. The quartets are stored sequentially, 
       i.e.\ the array elements read  $a_1, b_1, c_1, d_1, a_2, b_2, c_2, d_2, \ldots$, to reduce the memory swap when reading/writing the indices.
 \item A permutation number $-8 \le p \le 7$ in a one-dimension integer 1-byte array. As explained below, this index is used to reconstruct rapidly the time-reversed 
       matrix elements of $\bar{h}_{abcd}$ with the indices in the right order.
\end{enumerate}
With that, each matrix element takes 13 bytes of memory. The number of matrix elements itself is stored as a single 8-byte integer.

One could argue that it would be preferable to store the matrix elements as double precision 8-byte real numbers. While it is true to some extent, our tests
show that the difference in energy resulting from the two choices is below the keV level, i.e.\ much below the accuracy of any many-body method or interaction available.
In addition, the matrix elements are never used to perform sensitive numerical operations. Indeed, they are only used in the program in the calculations of the fields
where they are multiplied by the matrix elements of the one-body densities. 
Nevertheless, the code can be easily switched to a double precision storage of the matrix elements by setting "\texttt{rH2 = real64}"
in the file \texttt{module\_constants.f90} and in that case, each matrix element will take 17 bytes of memory. 

The vast majority of two-body matrix elements in the SHO basis will vanish due to the symmetries of the Hamiltonian such as the rotational and parity invariances or the conservation
of the number of neutrons and protons. These symmetries translate into the relations
\begin{subequations}
\begin{align}
 (-1)^{l_a + l_b} &= (-1)^{l_c + l_d}    \\
 m_{j_a} + m_{j_b} &=  m_{j_c}  + m_{j_d}  \\
 m_{t_a} + m_{t_b} &=  m_{t_c}  + m_{t_d}     
\end{align}
\end{subequations}
that can be used to reduce the loops over the indices when decoupling the $J$-scheme matrix elements. 

In addition, it is possible to reduce the number of matrix elements to be stored by considering their antisymmetry under the exchange of two particles, in the bra or the ket, and their 
symmetry\footnote{The matrix elements are considered real.} under the exchange of the bra and the ket. More precisely, we have 
\begin{subequations}
\begin{align}
 \bar{h}_{abcd} &=          -  \bar{h}_{abdc}  ,  \\
                &= \phantom{-} \bar{h}_{badc}  ,  \\
                &=          -  \bar{h}_{bacd}  ,  \\
                &= \phantom{-} \bar{h}_{cdab}  ,  \\
                &=          -  \bar{h}_{cdba}  ,  \\
                &= \phantom{-} \bar{h}_{dcba}  ,  \\
                &=          -  \bar{h}_{dcab}  ,     
\end{align}
\end{subequations}
such that only a subset of the matrix elements have to be stored. Indeed, the non-stored matrix elements can be reconstructed on the fly 
during the calculations of the fields by permuting the indices and multiplying by the appropriate phase factor. 
In the code, the single-particle basis states are labeled according to their
index $\in \llbracket 1, d_{\mathcal{M}} \rrbracket$. With that said, the matrix elements stored are such that 
\begin{equation}
\label{eq:indiset}
\begin{split}
 a &\in \llbracket 1, d_{\mathcal{M}} \rrbracket , \\
 c &\in \llbracket a, d_{\mathcal{M}} \rrbracket , \\
 d &\in \llbracket c+1, d_{\mathcal{M}} \rrbracket , \\
 b &\in \llbracket a+1, b_{\text{max}} \rrbracket \text{ with } \left\{ 
 \begin{array}{l}
   b_{\text{max}} = d \text{ if } c = a ,  \\
   b = d_{\mathcal{M}} \text{ otherwise.}
 \end{array} \right.
\end{split}
\end{equation}

Furthermore, it is possible to use the time-reversal invariance of $H$ to obtain
\begin{equation}
  \bar{h}_{abcd} = (-1)^{j_a + j_b + j_c + j_d} \bar{h}_{-a-b-c-d} .
\end{equation}
As the trial wave functions considered in \TAURUSvap~may break the time-reversal invariance, it is not possible to use this relation to its full
extent, i.e.\ reducing the sum over the single-particle states in the calculation of the fields.
Nevertheless, it allows us to store only the matrix elements that are such that 
\begin{equation}
m_{j_a} + m_{j_b} \ge 0 , 
\end{equation}
at the cost of reconstructing the 
time-reversed matrix elements when calculating the fields. Given the possible large number of matrix elements and the fact that the reconstruction
has to follow the rules of Eq.~\eqref{eq:indiset}, this takes a non negligeable amout of time.
Therefore, to reduce the time needed for the reconstruction, we store for each matrix element $\bar{h}_{abcd}$ an extra permutation index $p$
in a one-dimensional 1-byte integer array that can be used by the code to rapidly determine the correct order of indices and the appropriate phase
factor when reconstructing its time-reversed matrix element. 
This permutation index is calculated only once after the decoupling of the $J$-scheme matrix elements.

While efficient, this strategy is not sufficient by itself to tackle very large model spaces. Indeed, the memory required to store the full set of
matrix elements can rapidly exceed the RAM available on a single computer/node. To bypass this problem, the code can also share the matrix elements
among several MPI processes as described in Sec.~\ref{sec:para}.

%
%
\begin{acknowledgements}
We would like to thank M.~Frosini, J.~M.~Yao, A.~Tichai and V.~Somà for helping us benchmarking our code against their own solvers, and 
M.~Drissi and P.~Arthuis for testing the code and scripts.
We would also like to thank W.~Ryssens for helpful discussions on the Heavy-Ball method
and L.~M.~Robledo for fruitful discussions.
Finally, we would like to thank 
M.~Frosini for performing the transformation of the chiral interaction provided initially by R.~Roth.
T.~R.~R.~acknowledges the support from the Spanish Ministerio de Ciencia e Innovación under contract PGC2018-094583-B-I00.
This project has received funding from the European Union’s Horizon 2020 research and innovation programme under 
the Marie Skłodowska-Curie grant agreement No.~839847.
We acknowledge the computer resources and assistance provided by Centro de Computación Científica-Universidad Autónoma de Madrid (CCC-UAM).
\end{acknowledgements}
%
%
\appendix
%
%
\bibliographystyle{apsrev}

\bibliography{biblio}
%
%
\end{document}